\documentclass[12pt,a4paper,oneside,onecolumn]{book}
\usepackage[a4paper, margin=1in]{geometry}
\usepackage{pdfpages}

\usepackage{libertine}
\usepackage{courier}
\usepackage[hebrew,english]{babel}
\usepackage[longnamesfirst,square,numbers,comma,sort&compress]{natbib}
\usepackage{titlesec}
\usepackage{float}
\usepackage{graphicx}
\usepackage{subcaption}
\usepackage{amsmath}
\usepackage{mathpazo}
\usepackage{datetime}
\usepackage{setspace}
\usepackage{amsmath}
\usepackage{etoolbox}
\usepackage{rotating} % Include the rotating package
\usepackage{array}    % For advanced table features
\usepackage{soul}
\usepackage{xcolor}
\usepackage{longtable}
\usepackage{booktabs}
\usepackage{amssymb} % For \checkmark
\usepackage{multicol}
\usepackage{tabularx} % Add this to your preamble
\usepackage{multirow}

\usepackage{hyperref}
\hypersetup{colorlinks=true, linkcolor=blue, urlcolor=blue, citecolor=blue}

\sethlcolor{yellow}

\makeatletter
\renewcommand{\tagform@}[1]{,\ \textnormal{\normalfont(#1)}}
\makeatother

\titleformat{\chapter}[hang]{\bf\huge}{\thechapter}{2pc}{}
\title{HUJI M.Sc.~Thesis Sarah}

\begin{document}

\onehalfspacing
\includepdf[pages=1]{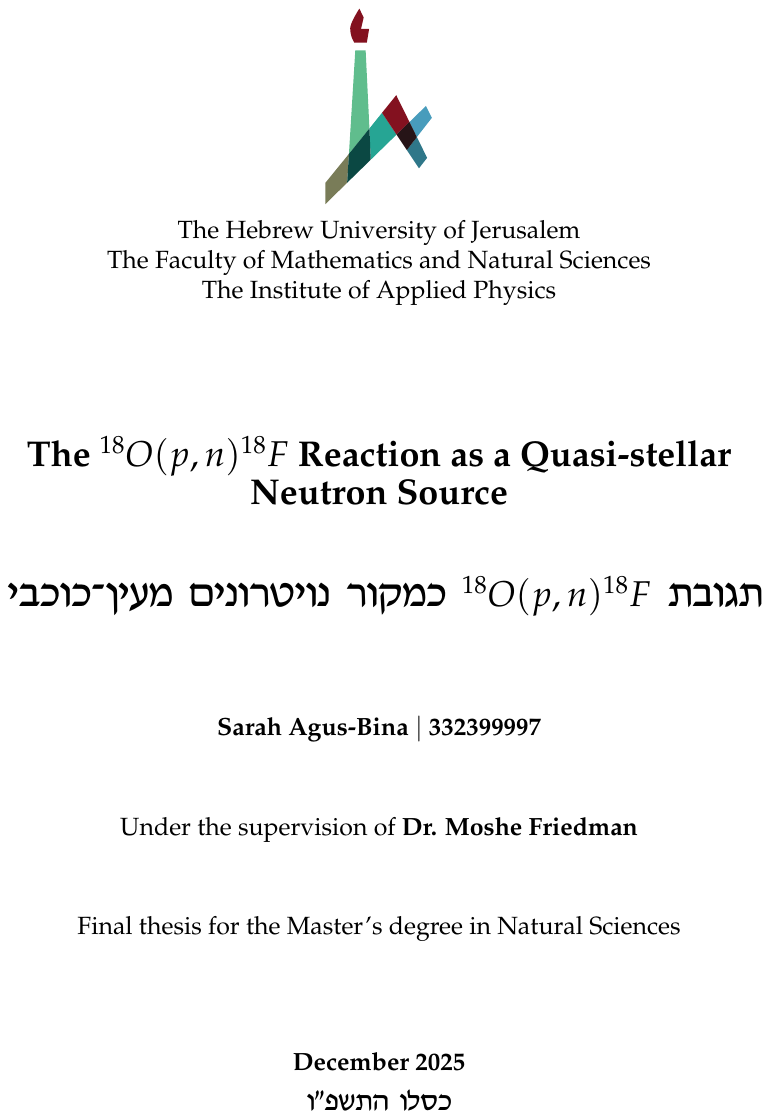}
%\includepdf[pages=1]{0.cover_heb.pdf}
%\includepdf[pages=1]{1.abstract_heb.pdf}
\frontmatter

\renewcommand{\baselinestretch}{1.5}

\chapter*{Acknowledgments}

I would like to express my deepest gratitude to all the individuals and institutions that have contributed to the completion of this thesis.

First and foremost, I would like to thank my supervisor, Dr. Moshe Friedman, for his exceptional guidance, patience, and unwavering dedication throughout my studies. His expertise, encouragement and mentorship have been instrumental in shaping my academic journey.

I am also grateful to all the past and present members of the laboratory for their collaboration, kindness, and the positive and inspiring atmosphere we shared. I appreciate each and every one of you for creating such a welcoming and productive environment, which made my time in the lab not only educational but also enjoyable.

A special thank you goes to the Physikalisch-Technische Bundesanstalt (PTB) for hosting us and guiding us through the experimental research. The opportunity to observe the experiment firsthand, gain insight into the facilities, and collaborate with the PTB team was highly enriching, and I am grateful for the knowledge and expertise they generously shared.

I am profoundly thankful to my husband, Ari, for standing by me with unwavering support and encouragement throughout this journey. His understanding and belief in me have been a constant source of strength. I am also deeply grateful to my parents for their lifelong support and for always encouraging my pursuit of knowledge. To my dear children, Moshe and Maayan: your energy, cuddles, joy and curiosity have been my daily inspiration and a constant reminder of what truly matters.

This thesis is the result of hard work and dedication, and I could not have completed it without the commitment, kindness, and insight of my family, friends, and colleagues. I am truly grateful to all who accompanied me on this path.
\chapter*{Abstract}

%There has to be an abstract written in English and a more detailed abstract written in Hebrew.

The slow neutron capture process (s-process) in asymptotic giant branch (AGB) stars produces elements with atomic mass numbers $A\gtrsim60$ through successive neutron captures and beta decays. In stellar environments where the s-process occurs, neutrons quickly thermalize, adopting a Maxwell-Boltzmann energy distribution determined by the local temperature independent of their production via $(\alpha,n)$ reactions. Laboratory experiments reproduce this Maxwell-Boltzmann neutron energy spectrum using $(p,n)$ reactions. Specifically, the $^{7}$Li$(p,n)^{7}$Be reaction is commonly employed to measure s-process cross-sections at $kT\approx25$ keV. Expanding the range of reactions used for measuring neutron-induced s-process cross-sections can offer valuable insights into the s-process in AGB stars. One such reaction is $^{18}$O$(p,n)^{18}$F, which can be used to mimic s-process cross-sections, as its neutron energy distribution is similar to the thermal flux distribution at the stellar environment where the $^{13}$C$(\alpha,n)^{16}$O reaction ($kT=8$ keV) takes place.

% old version:
%The Maxwell-Boltzmann energy distribution of neutrons produced by $(\alpha,n)$ reactions in the s-process can be mimicked in laboratory settings using $(p,n)$ reactions with similar Maxwellian flux energies. Specifically, the $^{7}$Li$(p,n)^{7}$Be reaction is commonly employed to measure s-process cross-sections at $kT\approx25$ keV, corresponding to the thermal neutron energy at the stellar environments such as the $^{22}Ne(\alpha,n)^{25}Mg$ reaction ($kT=23$ keV). Expanding the range of reactions used for measuring neutron-induced s-process cross-sections can offer valuable insights into the s-process in AGB stars. 

Heil \textit{et al.} showed that the neutron energy spectrum emitted from the $^{18}$O$(p,n)^{18}$F reaction at proton energy of $E_p = 2582$ keV, close to the reaction threshold of 2574 keV, closely resembles a Maxwellian flux with $kT\approx5$ keV. This experiment was repeated at PTB and a computational tool, OxyGen, was created to calculate the expected neutron energy spectrum and angular distribution at a planned liquid water-based $^{18}$O target at SARAF. OxyGen was incorporated into Geant4 transport simulations, which were compared to experimental results. The analysis shows that the simulated neutron energy spectrum from OxyGen agrees reasonably well with both current and previous experimental data.

Overall, the findings of this thesis demonstrate that the OxyGen simulation provides a reliable prediction of the neutron energy spectrum for the $^{18}$O$(p,n)^{18}$F reaction and can be effectively used to plan and analyze future experiments at SARAF for different proton energies.
\includepdf[pages=1-2]{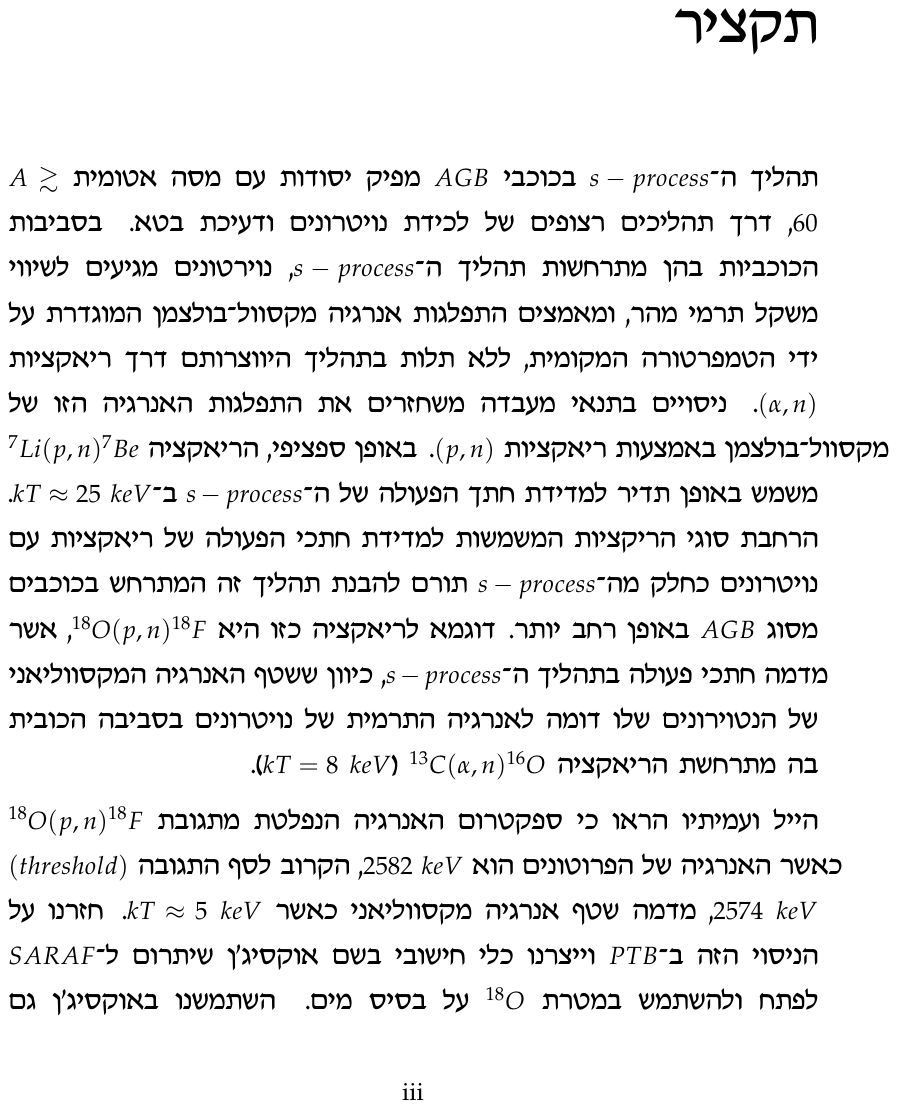}
\hypersetup{
    linkcolor=black, % Coloring TOC links black
}
\tableofcontents
\listoffigures
\listoftables
\hypersetup{
    linkcolor=blue, % Reverting to blue for the rest of the document
}

\mainmatter

\chapter{Introduction}
\label{chap:intro}

\section{Motivation}

The slow neutron capture process (s-process) is a sequence of nuclear reactions primarily occurring in asymptotic giant branch (AGB) stars, and is responsible for the production of elements with atomic mass numbers $A\gtrsim60$ \cite{Kappler2011}. In the s-process, a nucleus captures a neutron, forming an isotope with one higher atomic mass. This process continues until the newly formed isotope becomes unstable, at which point beta decay occurs, producing an element with the next higher atomic number.
 
In stellar environments where the s-process occurs, neutrons quickly reach thermal equilibrium, and their energies follow a Maxwell-Boltzmann distribution. In realistic s-process scenarios, a range of thermal energies must be considered, spanning approximately from 8 keV to 90 keV. The neutrons participating in the s-process are mainly produced by $(\alpha,n)$ reactions, such as $^{13}$C$(\alpha,n)^{16}$O operating at 8 keV and $^{22}
$Ne$(\alpha,n)^{25}$Mg operating at 23 keV.

The Maxwell-Boltzmann distributed energy spectrum of the neutrons produced by these stellar reactions can be created in the lab by $(p,n)$ reactions, corresponding to similar Maxwellian flux energies. Following the work of Ratynski and K\"appeler \cite{Ratynski1988}, the $^{7}$Li$(p,n)^{7}$Be reaction is widely used to measure s-process cross-sections at $kT\approx25$ keV \cite{Feinberg2012}, close to that of the stellar reaction $^{22}
$Ne$(\alpha,n)^{25}$Mg. Expanding the available experimental data by incorporating a broader range of reactions that emit neutrons can significantly enhance the study of the neutron induced s-process reactions. This, in turn, can provide valuable insights for further research on the s-process in AGB stars.

Heil \textit{et al.} showed that the neutron energy spectrum produced by the $^{18}$O$(p,n)^{18}$F reaction at a proton energy of $E_p = 2582$ keV, which is near the reaction threshold of 2574 keV, closely resembles a Maxwellian flux with $kT=5.1\pm0.1$ keV. This is similar to the thermal neutron energy in the stellar environment of the $^{13}$C$(\alpha,n)^{16}$O reaction, where $kT=8$ keV. Therefore, incorporating measurements of the neutron energy spectrum from the $^{18}$O$(p,n)^{18}$F reaction near the energy threshold provides an opportunity to extend the neutron energy spectrum beyond the existing data from the $^{7}$Li$(p,n)^{7}$Be reaction, thereby advancing the study of neutron-induced s-process reactions.

%The $^{13}C(\alpha,n)^{16}O$ reaction is a primary stellar neutron source that can later participate in reactions associated with the s-process. Therefore, gaining a more comprehensive understanding of the neutron energy spectrum emitted by the $^{18}O(p,n)^{18}F$ reaction will provide critical data for utilizing this reaction by the $^{18}O$ target as a quasi-stellar neutron source.

An enriched water (H$_2$$^{18}$O) target is under consideration at the Soreq Applied Research Accelerator Facility (SARAF), in order to produce an additional, robust quasi-stellar neutron source capable of withstanding SARAF's high-intensity proton beam. SARAF's wide-energy beam necessitates a computational tool to calculate the neutrons produced by the $^{18}$O$(p,n)^{18}$F reaction and incorporate this data into transport simulations. In order to develop a computational tool that accurately represents the neutron spectrum, it is essential to have detailed knowledge of the neutron spectrum at various angles near the reaction threshold. Therefore, Heil \textit{et al.}'s experiment was repeated to allow a thorough comparison of the experimental results with simulations.

\section{Previous works}

The foundation of this thesis is built upon the research conducted by Heil \textit{et al.} \cite{Heil2005}, whose experiment was essentially replicated in this study, with a detailed analysis of the results and a comparison to simulations.

Another key reference is the SimLiT simulation \cite{SimLiT}, which serves as the basis for the OxyGen simulation relevant for SARAF. The OxyGen simulation described in this work is a computational tool designed to model the neutron production from the $^{18}$O$(p,n)^{18}$F reaction, incorporating these results into transport simulations.

\section{Thesis outline}

The objective of this work is to study the neutron spectrum emitted from the $^{18}$O$(p,n)^{18}$F reaction. Chapter \ref{chap:theoreticalBackground} discusses the astrophysical motivation behind the research, as well as practical motivation for building a liquid H$_2$$^{18}$O target. Chapter \ref{chap:experimental_setup} details the experimental setup and measurement procedures. Chapter \ref{chap:data_analysis} explains the process of the data analysis, including the Geant4 transport simulation written for this purpose. Chapter \ref{chap:OxyGen} discusses the implementation of the OxyGen simulation and its uses for data analysis. Chapter \ref{chap:results_and_discussion} discusses the results of the analysis, as well as comparing them to Heil \textit{et al.}'s experimental results. 

\chapter{Theoretical Background}
\label{chap:theoreticalBackground}

%Much of the theoretical background is based on explanatory articles written by K\"appeler \textit{et. al}, \textit{The $s$ process: Nuclear physics, stellar models, and observations} \cite{Kappler2011} and Burbidge \textit{et. al}, \textit{Synthesis of the Elements in Stars} \cite{Burbidge1957}.

Stellar nucleosynthesis, the creation of chemical elements by nuclear reactions within stars, involves many nuclear processes. Some of these processes, such as proton capture and neutron capture, can produce elements with an atomic mass number of $A\gtrsim60$ \cite{Kappler2011,Burbidge1957}. This is significant because the fusion reactions leading to the creation of elements where $A\gtrsim60$ are endothermic; therefore, the fusion processes cannot substantially contribute to the creation of the heavy elements. 

%Other option for opening paragrpah:
%Neutron capture processes are crucial for the synthesis of elements with atomic mass numbers $A \gtrsim 60$, which typically occurs in later stages of stellar evolution. In contrast, fusion reactions that create lighter elements ($A \lesssim 60$) are exothermic and occur earlier in stellar lifecycles.

\section{Neutron capture}
\label{sec:neutronCapture}

%Proton capture is a nuclear reaction in which one or more protons collide with an atomic nucleus and form a heavier nucleus, a new element by definition (as elements differ by the amount of protons in their nucleus). As the electrical repelling force between protons is stronger with a decrease of the distance between the protons, the p-process is rare, occurring only in environments of high temperature and pressure. The isotopes of the chemical elements it creates are rare as well and lack neutrons. The extreme conditions where the p-process can occur can be found in supernovae or in the outer layers of massive stars. 

Neutron capture is the primary production mechanism for nuclei where $A\gtrsim60$ \cite{Krane}. Neutron capture is a nuclear reaction in which one or more neutrons collide with an atomic nucleus and form a heavier nucleus, potentially leading to a more unstable isotope that may undergo further transformations to other elements, such as beta decay or fission. In stellar environments, nucleosynthesis via neutron capture is divided into two main types of processes: slow neutron capture (s-process) and rapid neutron capture (r-process). 

In the r-process, the nucleus captures neutrons so quickly that it does not have time to undergo beta decay between captures, resulting in a highly unstable, neutron-rich nucleus. Once the neutron capture phase slows down, these unstable nuclei begin to undergo beta decay, transforming into stable nuclei and producing a broad range of heavy elements. This process typically occurs in environments with extremely high neutron densities so that neutrons can be absorbed at a rate much faster than the nuclei can decay, such as supernova explosions and neutron star mergers.

On the other hand, in the s-process the neutron capture occurs slower than the typical decay time, so the nucleus has time to transform into another element before capturing another neutron. The type of decay that occurs in the s-process is typically the beta decay, where a neutron emits an electron and antineutrino and transforms into a proton. In the s-process, a nucleus undergoes neutron capture and forms an isotope with one higher atomic mass. It continues to do so until the new isotope is unstable. Once this is the case, beta-decay occurs producing an element of the next higher atomic number. 

There are two types of s-processes: the main s-process and the weak s-process. These processes differ mainly in terms of neutron flux, temperatures involved, and types of elements they produce. The main s-process is the dominant type of s-process and it occurs in the cores of asymptotic giant branch (AGB) stars, where neutron fluxes are moderate and neutron captures are slow, due to the cooler temperatures and lower density in the outer layer of the AGB stars compared to supernovae and neutron star mergers (where the r-process can occur). The weak s-process occurs in different environments with lower neutron fluxes and higher temperatures, such as in the core of massive stars or in the helium-burning phases of low-mass stars, producing lighter elements and certain isotopes of heavier elements.

In stellar environments where the s-process occurs, neutrons quickly reach thermal equilibrium, and their energies follow a Maxwell-Boltzmann distribution. The thermal energies involved in the s-process typically range from approximately 8 keV to 90 keV. The neutrons participating in the s-process are mainly produced through $(\alpha,n)$ reactions. For example, neutrons produced by the $^{22}$Ne$(\alpha,n)^{25}$Mg reaction occur in He shell flashes of low-mass AGB stars, where they achieve a thermal equilibrium energy of around 23 keV. This enables the neutrons to participate in the main s-process. At the lower end of the thermal energy range involved in the s-process, the $^{13}$C$(\alpha,n)^{16}$O reaction takes place in the $^{13}$C pockets of thermally pulsing low-mass AGB stars, where the thermal equilibrium energy is approximately 8 keV.

\subsection{Maxwellian averaged cross-sections (MACS)}

Maxwellian averaged cross-sections (MACS) are used to model the neutron capture rates in stellar environments, where neutrons follow a Maxwell-Boltzmann distribution. In order to accurately model the rate of nuclear reactions in the star, the reaction cross-sections are averaged over the neutron energy spectrum. MACS values are defined by the reaction rate normalized by the most probable velocity of the Maxwell-Boltzmann distribution \cite{Elizabeth2024}:

\begin{equation}
MACS = \frac{\left<\sigma v\right>}{v_T}=\frac{2}{\sqrt{\pi}}\frac{1}{(kT)^2}\int_0^\infty\sigma(E) E e^{-E/{kT}} dE
\end{equation}

where $k$ is the Boltzmann constant, $T$ is the temperature of the stellar environment, and $\sigma(E)$ is the energy-dependent capture cross-section. In the study of the s-process, MACS values provide a direct representation of the reaction rate in stars at a given temperature. Experimentally, MACS values can be determined through two primary techniques: neutron activation measurements and the neutron time of flight (TOF) method. Both techniques measure cross-sections then calculate the MACS by assuming specific temperatures corresponding to stellar conditions.

\begin{enumerate}

  \item \textbf{Neutron activation measurements:} This technique involves irradiating a sample with neutrons, causing the nuclei within the sample to render radioactive. The resulting radioactive decay emits gamma rays, which are measured using a gamma ray detector. From the measured activity and the known neutron flux, the neutron capture cross-section can be calculated. Accurate determination of the flux and knowledge of the neutron spectrum are essential for this method, as the experimental cross-section is interpreted based on the corresponding neutron flux. 
  
  \item \textbf{Neutron TOF technique:} This technique involves measuring the time it takes for neutrons to travel a known distance from the source to the detector. From this, the neutron energy can be determined using the basic relation: $E=\frac{1}{2}m\left(\frac{d}{t}\right)^2$, where $E$ is the neutron energy, $m$ is the neutron mass, $d$ is the distance traveled, and $t$ is the TOF. Once the neutron energies are determined, the reaction rates, such as neutron capture, can be measured by recording the number of interactions at each energy. These interaction rates are then used to derive the neutron capture cross-section as a function of energy.
  
\end{enumerate}

Once the cross-section is measured at various energies, the MACS is obtained by averaging the cross-section over the Maxwell-Boltzmann energy distribution at a specific temperature that corresponds to the thermal energy of neutrons in stellar environments.

\section{The $^{7}$Li$(p,n)^{7}$Be reaction as a neutron source}
\label{sec:LiReactionAsNeutronSource}

The energy spectrum of neutrons produced in the stellar environment where the thermal equilibrium energy is around 23-25 keV can be reproduced in the lab by projecting protons onto a $^7$Li target, inducing the $^{7}$Li$(p,n)^{7}$Be reaction producing neutrons. One of the main advantages of this reaction is the high neutron intensity, which makes it an effective source for measuring neutron capture cross-sections. 
%Additionally, the neutron energy spectrum of this reaction closely resembles a Maxwellian distribution at a thermal energy of $kT = 25$ keV, which is near the thermal energy of the stellar environment where the $^{22}$Ne$(\alpha,n)^{25}$Mg reaction occurs, around $kT = 23$ keV. 
Therefore, neutrons generated by the $^{7}$Li$(p,n)^{7}$Be reaction near its threshold are commonly employed to measure the cross-section of s-process nucleosynthesis reactions \cite{BeerKappler1980}. 
Activation measurements using the $^{7}$Li$(p,n)^{7}$Be reaction as a neutron source are used to determine the MACS corresponding to a thermal energy of $kT = 25$ keV \cite{Ratynski1988}.

Soreq Applied Research Accelerator Facility (SARAF) is a proton and deuteron linear accelerator located at Soreq Nuclear Research Center, Yavne, Israel. SARAF's uniqueness is its intense beam power, up to 5 mA. SARAF Phase I accelerated protons up to 4 MeV. It is currently going through an upgrade and is expected to deliver beam energies up to 40 MeV. \cite{SARAF}. Therefore, SARAF is a useful facility for producing neutron beams via nuclear reactions such as $^{7}$Li$(p,n)^{7}$Be. 

Because SARAF's beam has high intensity, a standard solid Li or LiF target is unable to withstand its beam power \cite{mastersthesis_moshe}. Therefore, a windowless forced-flow closed loop liquid-lithium target (LiLiT) has been built. LiLiT was used at SARAF to produce neutrons at the energy of approximately 30 keV via the $^{7}$Li$(p,n)^{7}$Be reaction. The combination of SARAF and LiLiT produced the most intense quasi-stellar neutron source in the world for s-process measurements.

SimLiT \cite{SimLiT,mastersthesis_moshe} is a C++ simulation tool created to calculate the energy and emission angle of neutrons from the $^{7}$Li$(p,n)^{7}$Be reaction in different parameter settings, while the proton energy is close to the threshold ($1.88-3.6$ MeV). It can be used as an input for pre-experiment calculations as well as data analysis.

However, the $^{7}$Li method has its limitations. The $^{7}$Li$(p,n)^{7}$Be reaction produces neutrons at a single thermal energy, corresponding to $kT = 25$ keV, which limits the ability to study reactions at a wider range of stellar temperatures. The established way to overcome this limitation is by extrapolations based on cross-section evaluations \cite{Tessler2015}. However, systematical uncertainties on those extrapolations are hard to assess.

\section{The $ ^{18}$O$(p,n)^{18}$F reaction as a neutron source}

While the $^{7}$Li$(p,n)^{7}$Be reaction provides neutrons at $kT = 25$ keV as mentioned in Sec. \ref{sec:LiReactionAsNeutronSource}, increasing the range of available experimental data by enhancing the variety of reactions that emit neutrons as sources can expand the study of the neutron energy spectrum, thereby providing valuable insights for further research on the s-process in AGB stars. The $^{18}$O$(p,n)^{18}$F reaction, producing neutrons at energies around $kT =8$ keV, addresses this need and can aid in studying reactions at lower stellar temperatures. By utilizing both the $^{7}$Li$(p,n)^{7}$Be reaction and the $^{18}$O$(p,n)^{18}$F reaction for activation measurements, more accurate evaluations of cross-sections and MACS values can be achieved. 

Measurement of the neutron energy spectrum produced by the $^{18}$O$(p,n)^{18}$F reaction near the energy threshold of $2574$ keV provides an additional quasi-stellar neutron source in addition to the spectrum produced by the $^{7}$Li$(p,n)^{7}$Be reaction. 
This extended energy range can be used for activation measurements at lower stellar energies, in order to achieve activation measurements closer to the thermal energy of $kT=8$ keV, allowing for further analysis of stellar reactions. In particular, one of the main stellar neutron sources involved in the s-process, $^{13}$C$(\alpha,n)^{16}$O, operates at a thermal energy of $kT=8$ keV.

Heil \textit{et al.} \cite{Heil2005} measured the neutron energy spectrum of the $^{18}$O$(p,n)^{18}$F reaction at a proton energy of $2582$ keV, $8$ keV above the reaction threshold ($2574$ keV). They found that the resulting neutron spectrum closely resembles a Maxwellian distribution at a thermal energy of $kT=5.1 \pm 0.1$ keV. 

Several activation measurements have been conducted using $^{18}O$ as a neutron source. For instance, Heil \textit{et al.} \cite{Heil2005} measured the MACS of $^{138}$Ba at a thermal energy of $kT = 5.1\pm 0.1$ keV. These results were obtained through activation measurements with the $^{18}$O$(p,n)^{18}$F reaction and were compared with data from TOF measurements \cite{Beer1997}, showing good agreement. Further activation measurements using $^{18}$O were performed on $^{175}$Lu and $^{197}$Au, with their respective MACS values calculated by Heil \textit{et al.} \cite{Heil2008}. Additionally, Winckler \textit{et al.} \cite{Winckler2006} determined the neutron capture cross-section of $^{139}$La at a thermal energy of $kT \approx 5$ keV using $^{18}$O, and the relevant MACS values were calculated. Finally, Uberseder \textit{et al.} \cite{Uberseder2017} measured the cross-section of the $^{23}$Na$(n,\gamma)^{24}$Na reaction by activating NaCl samples at two quasi-stellar neutron spectra, those produced by $^{18}$O and $^7$Li.

\subsection{SARAF and liquid H$_2$$^{18}$O target}

Inspired by the success of LiLiT mentioned in Sec. \ref{sec:LiReactionAsNeutronSource}, there is a plan to develop a liquid target based on H$_2$$^{18}$O \cite{RuimiTalk2024}. The idea of this initiative is to use the $^{18}$O$(p,n)^{18}$F reaction in order to produce an additional powerful quasi-stellar neutron source at around 8 keV, that can withstand the proton beam power at SARAF.

SARAF's wide-energy beam requires a computational tool to calculate the produced neutrons from the $^{18}$O$(p,n)^{18}$F reaction and use it as an input for transport simulations. This tool will be important for the design stages of the experiment, and essential for the analysis stages. In order to achieve this, the OxyGen was developed (presented in Chapter \ref{chap:OxyGen}), which simulates the $^{18}$O$(p,n)^{18}$F reaction at various proton energies, similar to SimLiT for the $^{7}$Li$(p,n)^{7}$Be reaction. Therefore, Heil \textit{et al.}'s experiment was repeated in order to validate and fine-tune OxyGen by comparing it to experimental results in a thorough manner.
\chapter{Experimental Setup}
\label{chap:experimental_setup}

An experiment was conducted at the Physikalisch-Technische Bundesanstalt (PTB) in Braunschweig, Germany, in order to determine the neutron energy distribution of the $^{18}$O$(p,n)^{18}$F reaction at a proton energy of $E_p\approx2582$ keV, similar to the experiment of Heil \textit{et al.} \cite{Heil2005}.

 \begin{figure}[ht]
    \centering
    \includegraphics[width=0.8\textwidth]{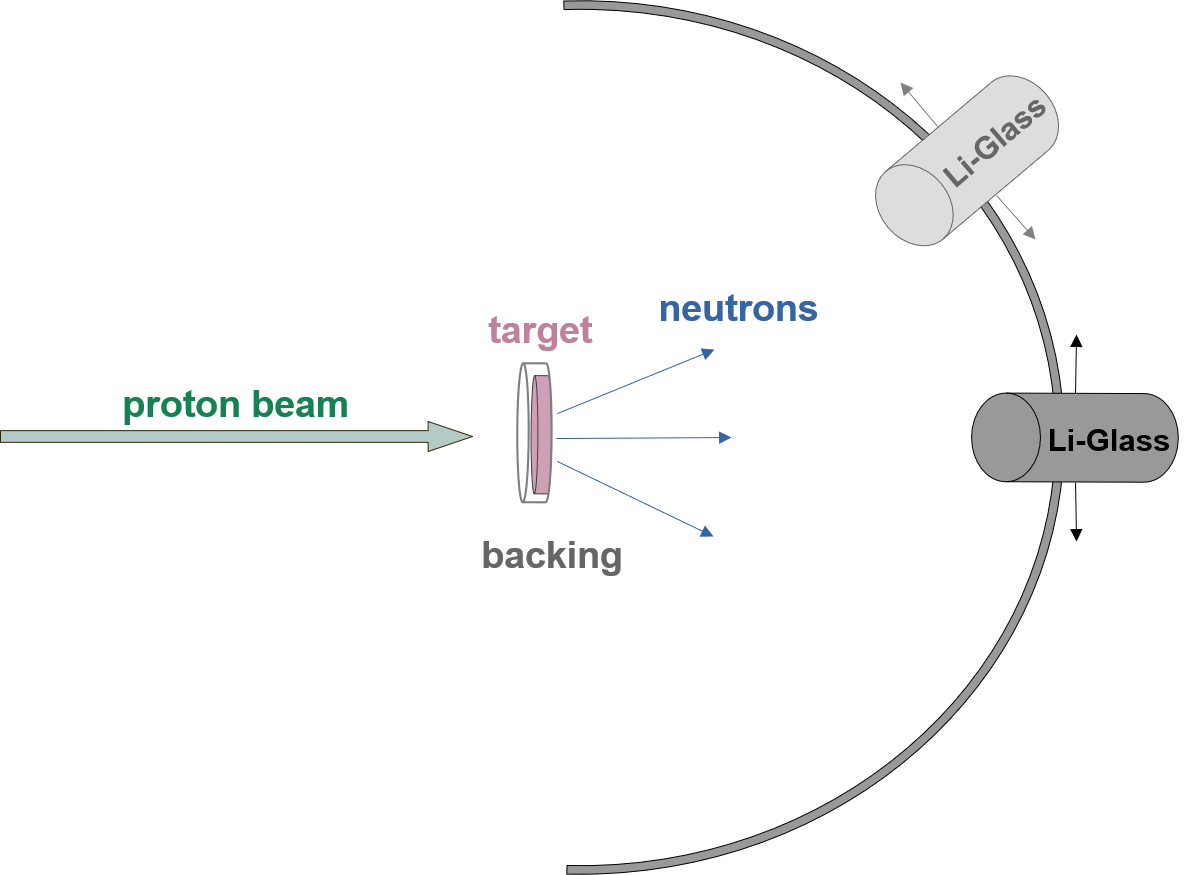}
    \caption{Experimental setup. A proton beam of $E_p=2580.5$ keV impinges on a thick Ta$_2^{18}$O$_5$ target with Ta backing. The proton beam induces the $^{18}$O$(p,n)$ reaction which serves as the neutron source. The neutrons are detected by a movable $^6$Li-glass detector, moving along a circular track at a distance of 20 cm. The proton beam is pulsed, and the neutron energy is determined by their time of flight (TOF). The experiment was repeated with the detector positioned at a 0\textdegree\ angle relative to the beam axis and up to 60\textdegree\ in 10\textdegree\ increments.}
    \label{fig:ExpSetup}
\end{figure}

\begin{figure}[ht]
    \centering
    \includegraphics[width=0.5\textwidth]{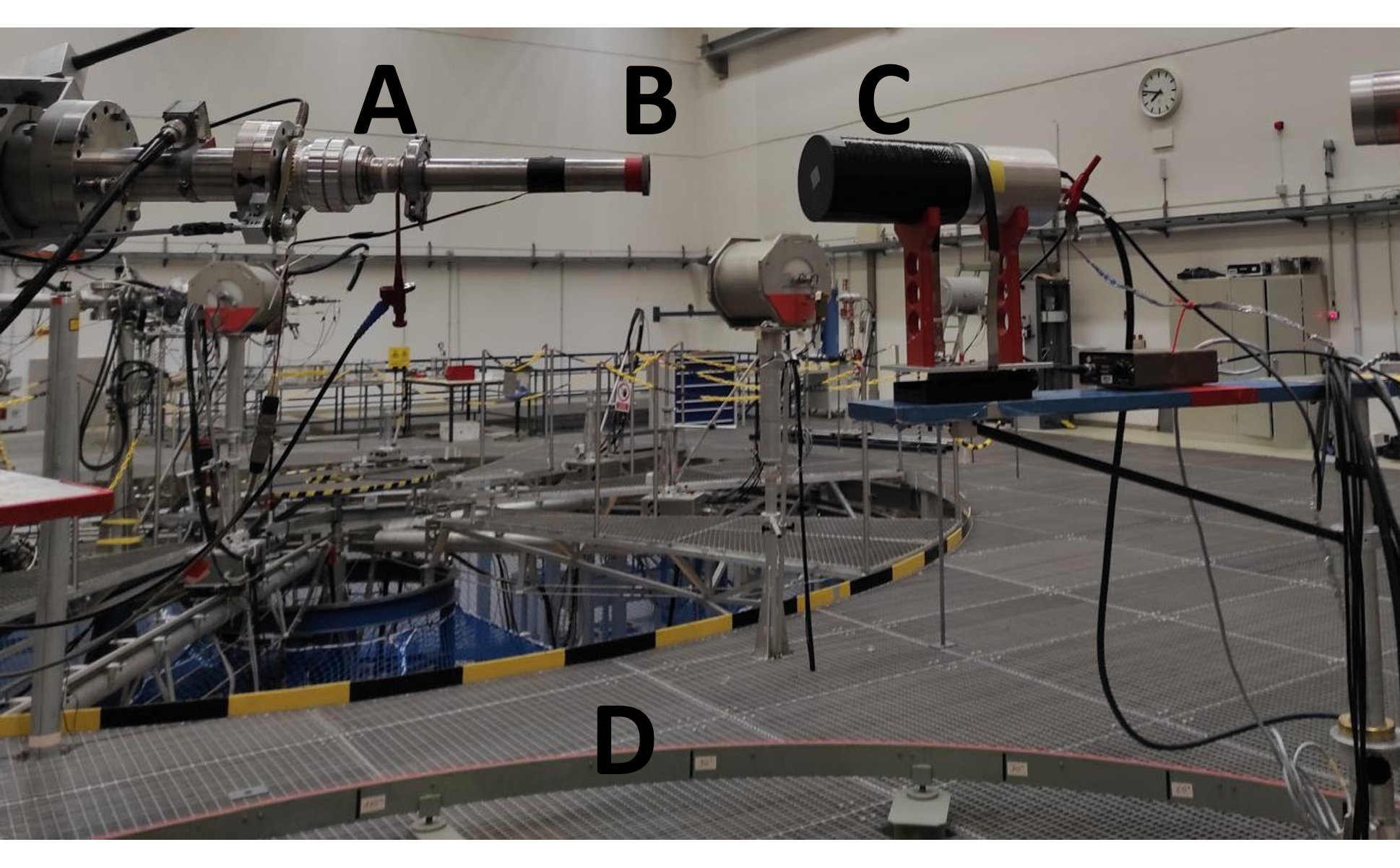}
    \caption{A picture of the experimental setup, including the proton beam (A), $^{18}$O target with backing (B), movable Li-glass detector (C) and the circular track on which the detector was placed (D).}
    \label{fig:ExpSetupImage}
\end{figure}

The neutron energy distribution of the $ ^{18}$O$(p,n)^{18}$F reaction was determined by a time of flight (TOF) experiment using the setup sketched in Fig. \ref{fig:ExpSetup}, which represents the configuration used at PIAF-PTB, as shown in Fig. \ref{fig:ExpSetupImage}. A pulsed proton beam was directed onto an $^{18}$O target with tantalum backing, releasing gamma rays and neutrons via the $^{18}$O$(p,\gamma)^{19}$F, $^{18}$O$(p,p^{\prime})^{19}$F, and $^{18}$O$(p,n)^{18}$F reactions. Because the proton energies are only slightly higher than the reaction thresholds, all neutrons were emitted in the forward direction. The neutrons (and gammas) were detected by the movable $^6$Li-glass detector. 

The target was fabricated by oxidizing a 0.5-mm-thick Ta disc in an $^{18}$O environment. The oxide layer thickness was measured by scanning the 334-keV resonance of the $^{18}$O$(p,\gamma)$ reaction to be 308 $\mu$g/cm$^2$, sufficient to slow down the protons below the $(p,n)$ energy threshold of 2574.8 keV. 

\section{The proton beam}
\label{sec:protonBeam}

The pulsed proton beam was provided by the PIAF-PTB accelerator operated at an energy of $2580.5$ keV with an energy width of $0.2$ keV. The actual beam energy, which was slightly below the requested one, was determined after the experiment as detailed in Sec. \ref{subsec:ProtonEnergies}. The repetition rate of the beam was $1.25$ MHz, and the average beam current was $0.5$ $\mu$A. The timing resolution was determined from the FWHM of the gamma peak, which is the combined timing resolution both from the pulse width and the detector response.

The proton beam energy was calibrated using three distinct methods. The first method involved scanning the $^{18}$O$(p,\gamma)$ resonance at 334 keV. The second method was a threshold scan, utilizing the known energy threshold of the reaction $^{18}$O$(p,n)^{18}$F at 2574.8 keV. In this approach, the beam energy was gradually increased starting below the threshold until neutrons were detected. The third method employed a TOF measurement at 0\textdegree; using basic kinematics, the TOF for the fastest neutron was calculated and then compared with the endpoint of the neutron TOF spectrum. These calibration methods were further validated through comparisons between the experimental data and simulations at different proton energies, as detailed in Sec. \ref{subsec:ProtonEnergies}.

\section{Li-glass detector}

\begin{figure}[ht]
    \centering
    \includegraphics[width=0.8\textwidth]{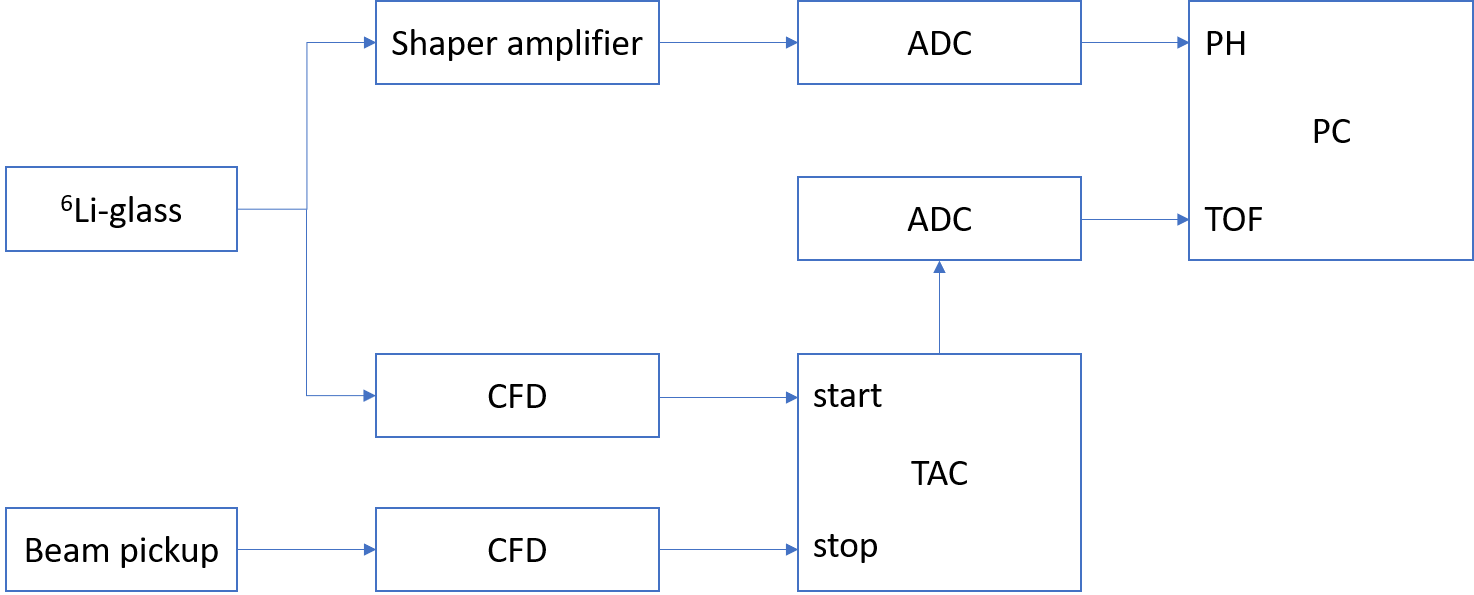}
    \caption{This sketch outlines the data acquisition scheme employed in the experiment.}
    \label{fig:DAQscheme}
\end{figure}

The neutron spectrum was determined by a movable $^6$Li-glass detector that was used to measure the TOF of neutrons at angles from $0^o$ to $60^o$ in $10^o$ increments. The detector was moved along a circular track. It consists of an enriched $^6$Li-glass which serves as the sensitive volume, coupled to a Photomultiplier Tube (PMT) that amplifies the signal for processing. A comprehensive description of the detector is presented in Sec. \ref{subsec:inputToG4}.

The neutrons are detected via the $^6$Li$(n,t)^4$He reaction. Upon detecting a neutron, the Li-glass releases an analog signal with a gamma-equivalent energy of 1.6 MeV. The analog signal from the PMT is split into two paths for simultaneous processing, as shown in Fig. \ref{fig:DAQscheme}. One path directs the signal through a shaper-amplifier, which transforms it into a standard Gaussian waveform. The amplitude of this waveform is proportional to the energy released in the detector. This signal is then digitized using an Analog-to-Digital Converter (ADC) and recorded as the pulse height (PH) in the data file.

The second signal is routed through a Constant Fraction Discriminator (CFD) and serves as the start signal for a Time-to-Analog Converter (TAC) (see Fig. \ref{fig:DAQscheme}). The stop signal for the TAC is generated from a beam pickup, which is also processed through a CFD. The TAC produces an analog signal proportional to the time difference between the detection event and the subsequent proton pulse. This TAC output is digitized by a second ADC and recorded as the TOF in the data file.

The final data file contains individual histograms for PH and TOF, as well as a 2D TOF-PH histogram.

The output of each ADC for a given signal is expressed in units of channels. In the case of the second signal proportional to the difference in time between the neutrons and protons detected, the calibration to units of nanoseconds was done by entering a periodic signal with a known and fixed time constant to the TAC.

During the experiment, every morning began with a measurement at $0^o$ in order to ensure the stability of the proton beam and data acquisition configuration.

\section{Experiment progression and adjustments}
\label{sec:ExpHistory}

The initial experiment was conducted in April 2022. However, the proton beam energy was too high relative to the target thickness, preventing the proton energy from decreasing below the reaction threshold. For accurate analysis, the proton energy distribution within the target needed to end below the reaction threshold, in order to take into account proton energies in the entire range from threshold to the nominal beam energy. Consequently, the experiment was repeated in June 2022 with adjustments to the proton beam energy.

To clarify, the neutron energy is determined by kinematic constraints for a given proton energy and angle, resulting in mono-energetic neutrons. The quasi-stellar distribution of these neutrons is achieved by two main factors: the integration of the solid angle and the variation in proton energy. When a proton with an initial energy of 2582 keV enters the Ta$_2$$^{18}$O$_5$ layer, it slows down and loses energy. As long as the proton's energy remains above the reaction threshold, it can interact with the $^{18}$O nuclei in the layer and produce a neutron via the $^{18}$O$(p,n)^{18}$F reaction.

If the Ta$_2$$^{18}$O$_5$ layer is sufficiently thick (on the order of a few hundred nanometers), protons can lose enough energy through interactions within the material, ensuring that (p,n) reactions occur over a range of proton energies, from the threshold to the nominal beam energy. This setup is referred to as a "thick target." In this configuration, increasing the target thickness beyond a certain point does not affect the $^{18}$O$(p,n)^{18}$F reaction, as no neutrons are produced below the reaction threshold.

Alternatively, a "thin target" setup, simply the thinnest target possible, can be used to produce mono-energetic neutrons. In this case, the target is thin enough to prevent significant slowing down of the protons, resulting in neutrons with well defined energy at any specific angle. During the first run of the experiment in April 2022, however, the target thickness was insufficient to ensure that the protons slowed down enough to reach this threshold, hence, the proton energy distribution was not that of a thick target. Furthermore, the target was not thin enough to be considered a "thin target" producing neutrons with well defined energy. As a result, the experiment produced a spectrum of neutron energies, which was a characteristic of a "thick target" setup, but did not meet the necessary condition of protons beginning beneath the reaction threshold for accurate analysis. To address this issue and obtain results more feasible for analysis, the experiment was repeated with a reduced proton beam energy in June 2022.

\section{Experiment information}

\begin{table}[h!]
\centering
%\begin{tabular}{|c|c|}
\begin{tabular}{|c|c|c|}
\hline

%\textbf{\Large Angle} & \textbf{\Large Date} \\ \hline
\textbf{Angle} & \textbf{Date} & \textbf{Total charge (mC)} \\ \hline

\multirow {5}{*}{0\textdegree} & 20.06.2022 & 1.43 , 4.31 \\ \cline{2-3}
& 21.06.2022  & 3.57 \\ \cline{2-3}
& 22.06.2022  & 2.73 \\ \cline{2-3}
& 23.06.2022  & 3.12 \\ \cline{2-3}
& 24.06.2022 & 2.15 \\ \hline
%\multirow {5}{*}{0\textdegree} & 20.06.2022 \\ \cline{2-2}
%& 21.06.2022  \\ \cline{2-2}
%& 22.06.2022  \\ \cline{2-2}
%& 23.06.2022  \\ \cline{2-2}
%& 24.06.2022 \\ \hline

\multirow {2}{*}{10\textdegree} & 20.06.2022 & \multirow{2}{*}{9.86} \\ \cline{2-2}
& 21.06.2022 &  \\ \hline

20\textdegree & 21.06.2022  & 12.21 \\ \hline

\multirow {2}{*}{30\textdegree} & 21.06.2022 & \multirow {2}{*}{13.38} \\ \cline{2-2}
& 22.06.2022 & \\ \hline

40\textdegree & 22.06.2022  & 15.35 \\ \hline

\multirow {2}{*}{50\textdegree} & 22.06.2022 & \multirow {2}{*}{26.79} \\ \cline{2-2}
& 23.06.2022 & \\ \hline

\multirow {2}{*}{60\textdegree} & 23.06.2022 & \multirow {2}{*}{21.05} \\ \cline{2-2}
& 24.06.2022 & \\ \hline

\end{tabular}
\caption{Measurement dates for each detector angle, along with the total collected charge for each measurement. Two measurements were taken at 0\textdegree\ on 20.06.2022, hence the two charge values.}

\label{tab:measurementDatesPerAngle}
\end{table}

The experiment was executed for each angle according to the dates described in Table \ref{tab:measurementDatesPerAngle}. As shown, a measurement of 0\textdegree\ was taken each day in order to ensure the stability of the beam.

\chapter{Data Analysis}
\label{chap:data_analysis}

\section{Overview}
\label{sec:Overview}

The data analysis for this experiment involved applying a cut on the neutron Pulse Height (PH) to remove most of the gamma background, extracting the neutron time of flight (TOF) spectrum, subtracting background contributions from uncorrelated neutrons and remaining gamma detections, and converting the TOF data into energy. This conversion relied on the detector response as calculated by the Geant4 transport simulation, followed by an unfolding method based on Bayes' theorem. A correction for the solid angle was then applied. Finally, the results were compared with OxyGen to assess its reliability.

\section{Raw data and background treatment}

\begin{figure}[ht]
    \centering
    \includegraphics[width=0.8\textwidth]{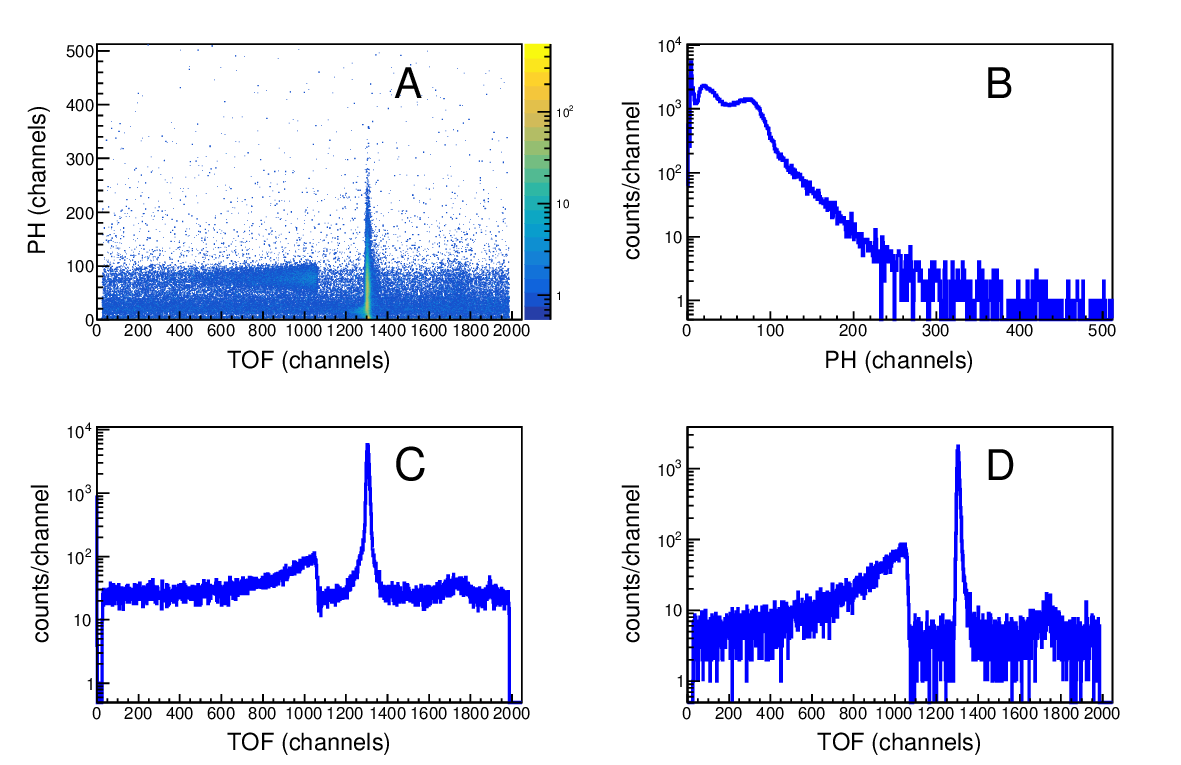}
    \caption{The pulse height (PH) and time of flight (TOF) for $0^\circ$ are shown both as a 2D histogram and as projections onto each axis, with the counts displayed on a logarithmic scale (z-axis for panel A and y-axis for panels B-D). The 2D histogram of PH versus TOF is presented in panel A, the 1D projection onto PH is shown in panel B, and the 1D projection onto TOF in panel C. Panel D shows the TOF projection restricted to events with PH values between 60 and 110 channels.}
    \label{fig:rawTOFvsPH}
\end{figure}

The raw experimental spectrum at $0^\circ$ is shown in Fig. \ref{fig:rawTOFvsPH}. This figure presents a 2D histogram of PH versus TOF (on a logarithmic scale), along with 1D histograms of both PH and TOF. Additionally, it includes the projection of the 2D histogram onto the TOF axis, after applying a cut on the PH range of $60\text{--}110$ channels.

When a neutron undergoes the $^6$Li$(n,t)^4$He reaction, the induced PH signal in the detector corresponds to the response of a 1.6 MeV gamma. Since the neutron's kinetic energy is negligible in comparison (energies up to $\sim$25 keV), the PH spectrum exhibits a distinct peak in the range of $\sim60\text{--}110$ channels. This peak is used as a selection criterion to suppress the overwhelming low-energy gamma background in the TOF domain. The primary sources of this gamma background are the nuclear reactions $^{18}$O$(p,\gamma)^{19}$F and $^{18}$O$(p,p\prime)^{18}$O.

Despite this selection cut, a distinct gamma peak remains visible in the TOF spectrum, which can be utilized for time calibration.
Note that the TOF spectra are inverted, meaning that higher channel numbers correspond to shorter TOF values. This inversion occurs because the detector signal was used as the stop signal, while the beam pickup served as the start. This approach was chosen to reduce data load, as the majority of proton bunches do not result in a detection event in the $^6$Li-glass.

\begin{figure}[ht]
    \centering
    \includegraphics[width=0.8\textwidth]{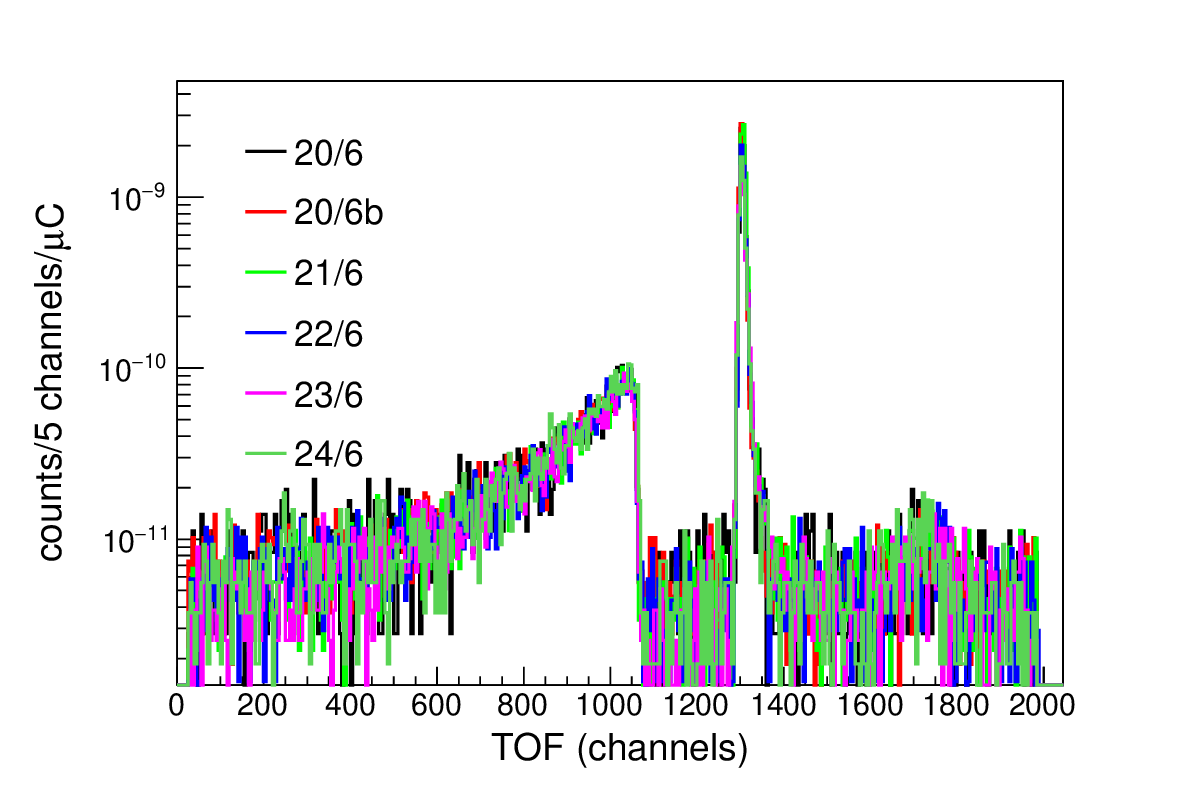}
    \caption{Daily TOF spectra measured at $0^\circ$ throughout the experiment, demonstrating the stability of the proton beam. The spectra were obtained after applying the PH cut of 60-110 channels, consistent with the extraction shown in Fig. \ref{fig:rawTOFvsPH}D.}
    \label{fig:tofStability}
\end{figure}

In order to verify beam stability and target integrity throughout the experiment, the spectrum at $0^\circ$ was taken each morning. The TOF spectrum, after applying the PH cut and normalizing to the total charge, is compared in Fig. \ref{fig:tofStability}. This comparison shows excellent beam stability and no apparent degradation in the target.

An additional background subtraction was done to account for uncorrelated events. This background was estimated by fitting the background in the region between the gamma peak and the neutron cutoff in the TOF spectrum.

The remaining clean TOF spectra are uncalibrated. The channel to ns time constant was determined by applying a known periodic signal to be $\eta=0.4035$ ns/channel. The gamma peak was fitted to a Gaussian and its mean, $\mu$ was used to calibrate the time $t^{\prime}$ based on the uncalibrated time $t$:

\begin{equation}
    t^{\prime} = \eta(\mu-t)+t_0
\end{equation}
where $t_0 = \Delta x/c$ is the mean flight time of the gamma peak. Finally, the calibrated TOF histograms were normalized to the applied charge.

\begin{figure}[ht]
    \centering
    \includegraphics[width=0.8\textwidth]{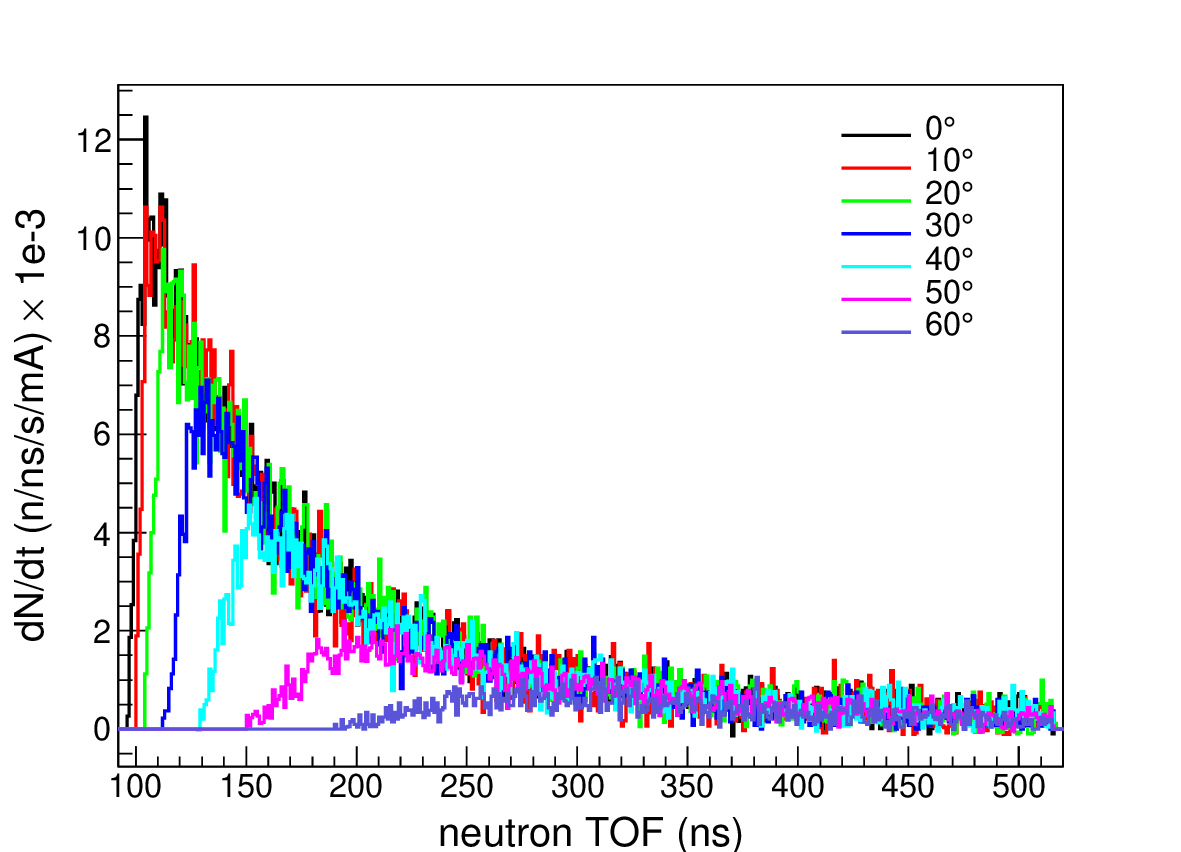}
    \caption{Experimental TOF spectra normalized to the charge for each angle. The height of the peak decreases with increasing angles. The background below the neutron cutoff is artificially removed to avoid numerical artifact in the Bayes' unfolding algorithm.}
    \label{fig:TOFallAngles}
\end{figure}

Fig. \ref{fig:TOFallAngles} shows the final TOF histograms for each angle of detection, after having completed the steps explained above. There is a systematic decrease in the integral with increasing angle, because the solid angle has not yet been accounted for.

\section{Geant4 transport simulation}
\label{sec:G4}

As mentioned in the overview of this chapter (Sec. \ref{sec:Overview}), a transport simulation was developed using Geant4 to model the detector's response and efficiency. This simulation accounts for the detector's materials, geometry, and placement to ensure accurate results.

The simulation takes neutrons with a defined momentum distribution as input and tracks their interactions within the detector. It calculates key parameters such as TOF, energy deposition, and detection efficiency. By varying the input energy distribution, the simulation generates a response matrix that can be used to unfold the experimental TOF spectrum into the corresponding neutron energy spectrum.

\subsection{Geometry}

 \begin{figure}[ht]
    \centering
    \includegraphics[width=0.5\textwidth]{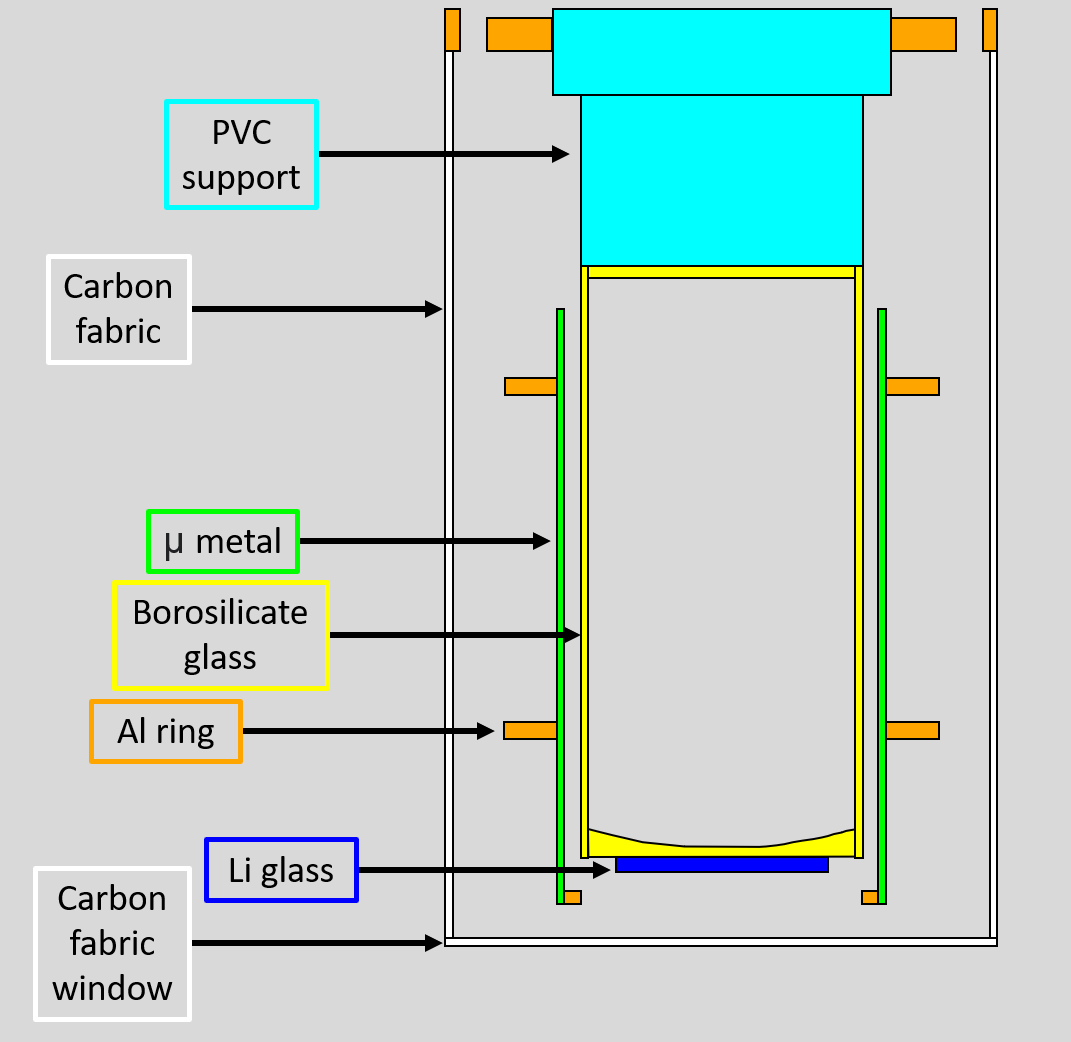}
    \caption{$^6$Li-glass detector geometry. See Table \ref{tab:materialDescriptionPTB} for dimensions and chemical compositions.}
    \label{fig:geometryFromPTB}
\end{figure}

\begin{table}[h!]
\centering
\begin{tabular}{|c|c|c|}
\hline
\textbf{Material name} & \textbf{Composition (\%atom)} & \textbf{Placement (cm)} \\ \hline

\multirow {6}{*}{Li-glass} & $^6$Li: 23.217 & \multirow {6}{*}{\parbox{2cm}{R=1.945 \\ T=0.285 \\ D=1.47}} \\ \cline{2-2}
& $^7$Li: 1.222 & \\ \cline{2-2}
& $^{16}$O: 54.371 & \\ \cline{2-2}
& Si: 20.727 & \\ \cline{2-2}
& $^{140}$Ce: 0.412 & \\ \cline{2-2}
& $^{142}$Ce: 0.052 & \\ \hline

\multirow {7}{*}{Borosilicate glass} & $^{10}$B: 0.702 & \multirow {7}{*}{\parbox{2cm}{R=2.6 \\ r = 2.5 \\ L=10.8375 \\ D=1.6125}} \\ \cline{2-2}
& $^{11}$B: 2.826 & \\ \cline{2-2}
& $^{16}$O: 67.267 & \\ \cline{2-2}
& $^{23}$Na: 1.673 & \\ \cline{2-2}
& $^{27}$Al: 1.156 & \\ \cline{2-2}
& Si: 26.166 & \\ \cline{2-2}
& K: 0.209 & \\ \hline

\multirow {6}{*}{$\mu$ metal} & $^{54}$Fe: 1.126 & \multirow {6}{*}{\parbox{2cm}{R=3.0 \\ r=2.9 \\ L=11.0 \\ D=0.705}} \\ \cline{2-2}
& $^{56}$Fe: 17.815 & \\ \cline{2-2}
& $^{57}$Fe: 0.427 & \\ \cline{2-2}
& $^{58}$Ni: 56.714 & \\ \cline{2-2}
& $^{60}$Ni: 21.090 & \\ \cline{2-2}
& $^{62}$Ni: 2.828 & \\ \hline

\multirow {4}{*}{Carbon fabric} & $^{1}$H: 43.84 & \multirow {4}{*}{\parbox{2cm}{R=5.050 \\ r=4.945 \\ L=16.4}} \\ \cline{2-2}
& C: 50.21 & \\ \cline{2-2}
& $^{14}$N: 2.23 & \\ \cline{2-2}
& $^{16}$O: 3.72 & \\ \hline

\multirow {4}{*}{Carbon fabric window} & $^{1}$H: 43.84 & \multirow {4}{*}{T=0.05} \\ \cline{2-2}
& C: 50.21 & \\ \cline{2-2}
& $^{14}$N: 2.23 & \\ \cline{2-2}
& $^{16}$O: 3.72 & \\ \hline

\multirow {2}{*}{Al ring} & $^{27}$Al: 97 & \multirow {2}{*}{\parbox{4cm}{R=2.9 ; r=2.6 \\ T=0.3 ; D=0.705}} \\ \cline{2-2}
& Mg: 3 & \\ \hline

\multirow {3}{*}{PVC support} & $^1$H: 50 & \multirow {3}{*}{\parbox{2cm}{R=2.6 \\ T=3.15 \\ D=12.45}} \\ \cline{2-2}
& C: 33.3 & \\ \cline{2-2}
& Cl: 16.7 & \\ \hline

\end{tabular}
\caption{Composition and placement of materials within the Li-glass detector provided by PTB. This is a description of the geometry presented in Fig. \ref{fig:geometryFromPTB}. 
The material composition is given as the atomic percentage of each isotope of the material. In the "Placement" column, R denotes the outer radius, r the inner radius, T the thickness, L the total length, and D the distance from the object's midpoint to the front of the entrance window.}
\label{tab:materialDescriptionPTB}
\end{table}

 \begin{figure}[ht]
    \centering
    \includegraphics[width=0.5\textwidth]{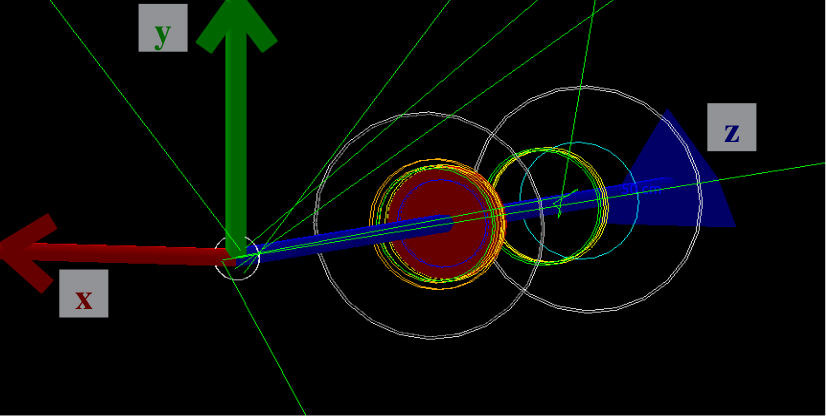}
    \caption{A graphical display of the geometry setup of the detector in the Geant4 simulation, showing example neutron trajectories. The backing of the target is included as well. Each color represents a different material.}
    \label{fig:G4geo}
\end{figure}
 
A detailed description of the detector geometry is provided in Fig. \ref{fig:geometryFromPTB} and Table \ref{tab:materialDescriptionPTB}\footnote{R. Nolte, Private Communication.}. All relevant materials and components were incorporated into the simulation. The detector's angle relative to the beam axis is configurable as a parameter, allowing for comparisons with TOF spectra from the OxyGen neutron source. The tantalum (Ta) backing can also be included in the geometry, and the detector's supporting arm is omitted.

Fig. \ref{fig:G4geo} presents a visual representation of the detector setup within the Geant4 simulation, along with example neutron trajectories to illustrate possible interaction paths.

\subsection{Simulation input}
\label{subsec:inputToG4}

The input of the Geant4 transport simulation is a neutron energy spectrum and angular distribution, and it calculates the efficiency, response matrix, and TOF spectrum. The neutron energy spectrum that is entered as the input is flexible. However, we chose to introduce a uniform distribution of neutrons between (0,30) keV as input to avoid introducing bias to the calculated efficiency and response matrix. When this was used as the input, the target backing was not taken into account, and the detector was fixed at a $0^{\circ}$. The angular distribution was uniform within the range of $\sim0^{\circ}-14^{\circ}$ to include only the neutrons that enter the detector, ensuring that no neutrons outside this range were considered for the efficiency calculation.

\subsection{Efficiency}

\begin{figure}[ht]
    \centering
    \includegraphics[width=0.6\textwidth]{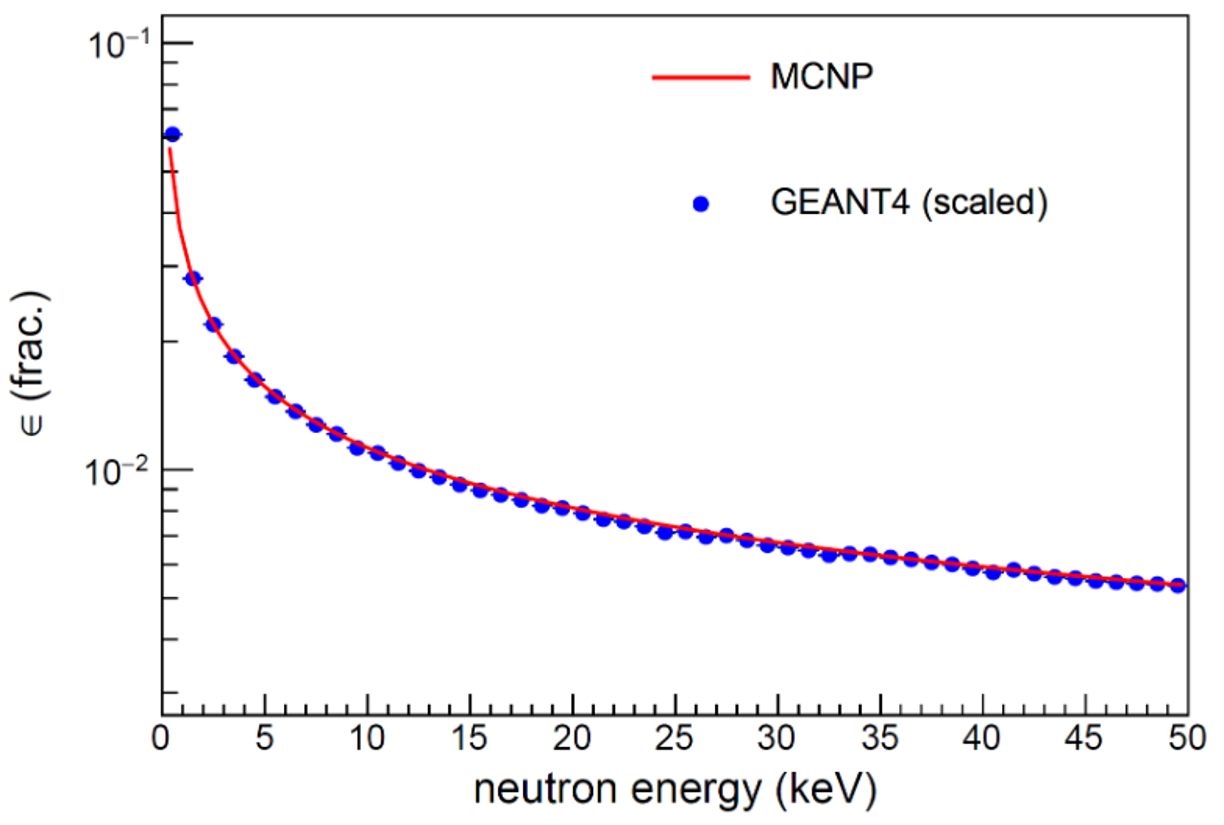}
    \caption{Detector efficiency calculations. The red curve is an MCNP simulation done by Ralf Nolte (PTB), and the blue circles represent our Geant4 transport simulation. We applied a scaling factor to the Geant4 efficiency curve to account for the unknown $\Delta\Omega$ used in the MCNP calculation (see text for explanation).}
    \label{fig:efficiencyComparison}
\end{figure}

The efficiency, $\epsilon$, is defined as the ratio between the number of detected neutrons, $N_C$, and the number of neutrons passing through a specific solid angle, $N_{\Delta\Omega}$, and it is calculated as a function of the neutron energy, $E$:
\begin{equation}
    \epsilon(E) = \frac{N_{C}(E)}{N_{\Delta\Omega}(E)}.
\end{equation}
The exact definition of $\Delta\Omega$ is somewhat arbitrary. Naturally, $\Delta\Omega$ will be the solid angle covered by the detector, which differs with the specifics of each experiment. In our case, the sensitive volume of the detector covers approximately half of the total solid angle of the detector, and neutron scattering off the different components of the case has a non-negligible probability of ending up in the sensitive volume. Here, we define $\Delta\Omega$ as the solid angle covered by the external carbon fabric window (see Fig. \ref{fig:geometryFromPTB}). An independent efficiency curve for the same detector was calculated in the past using MCNP\footnote{R. Nolte, Private Communication.}. Fig. \ref{fig:efficiencyComparison} shows a good agreement between the calculations. We do not have information regarding the exact definition of $\Delta\Omega$ in the MCNP calculation, hence we scaled our calculation to compare the energy dependency alone. 

In practice, the analysis was done by inverting the response matrix, as described below. The efficiency is incorporated into the response matrix, and was not used on its own.

\subsection{Response matrix}

 \begin{figure}[ht]
    \centering
    \includegraphics[width=0.7\textwidth]{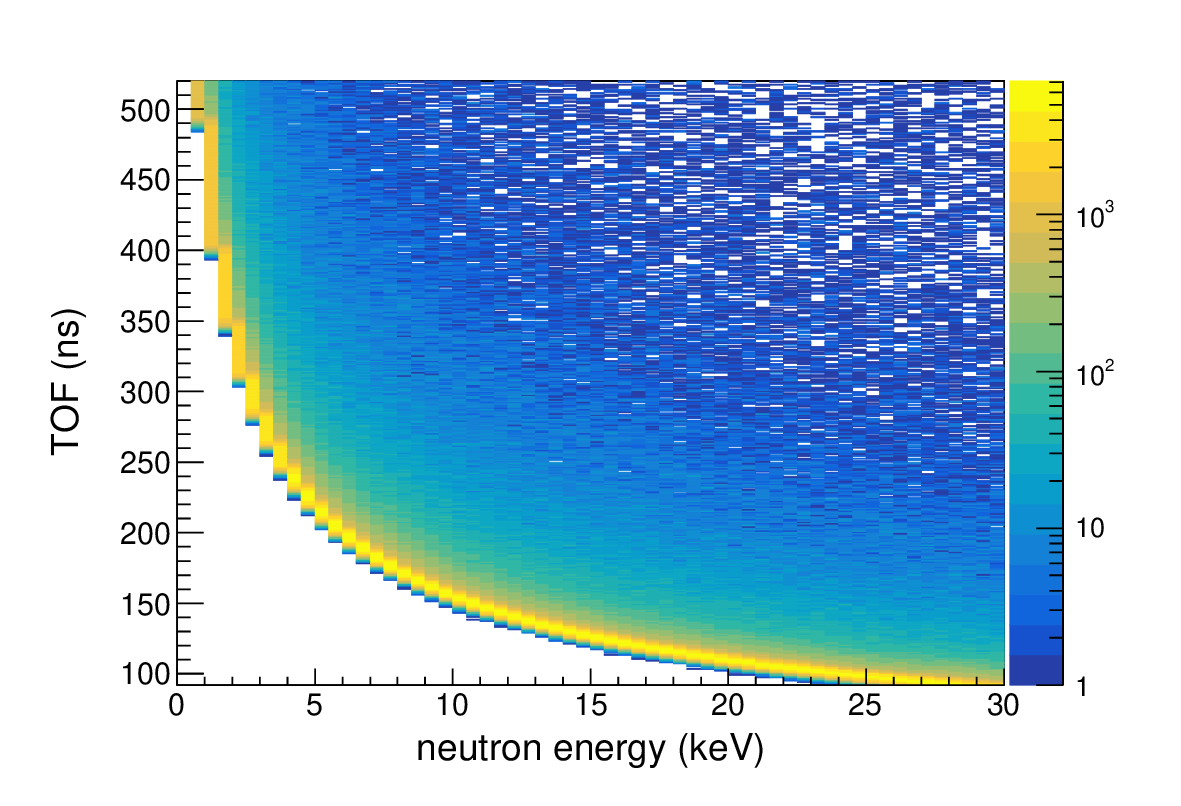}
    \caption{Detector response matrix presented in 2D histogram form. It shows the simulated TOF vs energy for the transport simulation using uniformly distributed neutrons in the energy range of (0,30) keV as the input. The z axis is in log scale. The strong yellow band roughly follows the relation $E=\frac{1}{2}mv^2$, and the tails to longer times are due to neutron scattering in the detector.}
    \label{fig:responseMatrix}
\end{figure}

The conversion of TOF to neutron energy is typically performed using the nonrelativistic equation:  
\begin{equation}
    E_n = \frac{1}{2} m_n \left(\frac{x}{t}\right)^2.
\end{equation}  
Here, $E_n$ and $m_n$ denote the neutron energy and mass, respectively, while $x$ represents the mean flight path and $t$ the TOF. The relative uncertainty in energy is given by:  
\begin{equation}
    \frac{\Delta E_n}{E_n} = \sqrt{\left(\frac{2\Delta x}{x}\right)^2 + \left(\frac{2\Delta t}{t}\right)^2}
\end{equation}  
However, this straightforward approach does not account for neutron scattering in the detector, which effectively increases the flight path $x$ in a non-trivial manner. This effect has been discussed in Ref. \cite{Feinberg2012} and is evident in Fig. \ref{fig:responseMatrix}. The figure presents the detailed response matrix, showing the calculated TOF distribution for each neutron energy in 1-keV bins.  

Ideally, the neutron energy distribution could be obtained by multiplying the inverted response matrix by the measured TOF distribution. Unfortunately, direct matrix inversion is not feasible, and instead, Bayes' unfolding method was employed.  

\section{Bayes' unfolding method}
\label{sec:BayesUnfoldingMethod}

D'Agostini \cite{DAgostini1995} proposed an unfolding method based on Bayes' theorem to overcome the difficulties of the matrix inversion method, namely its inapplicability with singular matrices, as is the case for the response matrix calculated by the Geant4 transport simulation. Looking at the TOF for a specific neutron energy, we obtain a distribution, and not a single TOF for a single neutron energy. Meaning, if we were to convert the TOF back to energy, we would receive a different distribution than the one we started with. This is the physical explanation of a singular matrix being non-invertible. The energy can be transformed to TOF but the TOF cannot be transformed into neutron energy spectrum terms. The unfolding method based on Bayes' theorem overcomes this issue.

As D'Agostini thoroughly explains \cite{DAgostini1995}, Bayes' theorem states:

\begin{equation}
P(C_i|E) = \frac{P(E|C_i)P(C_i)}{ \Sigma_{l=1}^{n_c} P(E|C_l)P(C_l)}
\label{eq:BayesPre}
\end{equation}

for an event $E$ and different causes $C_i$, while the index of different possible causes is given by the notation $i=1,2,...,n_c$. The equation determines the probability of cause $C_i$ given the event $E$, $P(C_i|E)$, in terms of the probability of event $E$ given the cause $C_i$, $P(E|C_i)$, and the probability of cause $C_i$, $P(C_i)$, divided by a normalization factor, which is the total probability for an event $E$. In more intuitive terms, given an effect $E$, the probability of a certain cause being the source of the effect is determined by the probability of the effect given the cause multiplied by the probability of the cause. 

For example, considering a baby crying as the effect \( E \), there are many possible causes for this. To name a few: the baby is hungry, the baby is tired, or the baby needs a diaper change. We can ask what the probability is that hunger is the cause of the baby's crying. The answer depends on two factors: the likelihood that a hungry baby cries, \( P(E \mid C_i) \), and the probability that the baby is hungry, \( P(C_i) \).  

\( P(C_i) \) likely depends on several factors, such as the time since the baby last ate, their age, and their typical eating schedule. We use our prior knowledge of \( P(C_i) \) to refine our estimate of the probability that the baby is crying due to hunger, \( P(C_i \mid E) \). This updated probability, \( P(C_i \mid E) \), becomes our new estimate for \( P(C_i) \). Since \( P(C_i) \) represents our initial belief before considering the evidence, it is called the \textit{prior}. After calculating \( P(C_i \mid E) \), this updated probability is referred to as the \textit{posterior}.

Implementing this theory in the experimental data analysis, the "causes" $C_i$ correspond to the neutron energies, while the "effects" $E_j$ represent the TOF data. For each cause $C_i$ (energy), a distribution of TOF values $E_j$ is expected. This relationship is essentially represented by the response matrix obtained from the transport simulation. In reality, we observe a TOF spectrum generated by an underlying distribution of causes. Assuming a uniform prior distribution, we can use Bayes' theorem to derive a posterior energy distribution:

\begin{equation}
    \hat{n}(C_i) = \sum_{j=1}^{N_E} n(E_j) P(C_i | E_j).
\end{equation}

Here, $n(E_j)$ is the number of neutrons detected at TOF $E_j$, and $\hat{n}(C_i)$ is our posterior estimate of the number of neutrons emitted at energy $C_i$, or, equivalently, our posterior estimate of the neutron energy spectrum. The term $P(E_j | C_i)$ represents the response matrix calculated using our Geant4 simulation. 

However, the quality of this neutron energy spectrum estimate depends on our prior assumption. There is no strong reason to assume that a uniform prior is accurate. Fortunately, there is a straightforward way to assess this assumption: we can apply the response matrix to the posterior distribution and compare the resulting TOF spectrum with the experimental data. If the match is insufficient by some criterion, we recognize that our result is not yet satisfactory. Nevertheless, the key advantage of this approach is that the posterior distribution is expected to be closer to the true energy distribution than the initial uniform prior. This observation suggests that we can iteratively refine the posterior until it meets our criteria. In this regard, this process can be viewed as an alternative to gradient descent rather than a conventional Bayesian analysis.

In practice, Bayes' method of unfolding is implemented in the RooUnfold package \cite{RooUnfold} and includes three main steps. The first step is determining the prior. In our case, we used a uniform neutron energy distribution between $(0,30)$ keV as the initial prior. The second step is calculating the posterior using Bayes' theorem. Third, a calculation of $\chi^2$ between the prior and posterior was performed in order to evaluate the difference. If the difference is above some arbitrary threshold, the posterior is used as the new prior and the process is repeated. When the process is complete, the neutron energy spectrum is considered extracted. 

It is important to highlight that Bayes' unfolding method inherently accounts for the efficiency correction through the use of the response matrix derived from the transport simulation. This matrix incorporates the efficiency by reflecting how events are distributed across the energy axis, including those where no TOF signal was recorded. As a result, the integral over the TOF axis for a given energy is not uniform across different energy values, despite the initial assumption of a uniform energy distribution. By using the response matrix, the efficiency effects are embedded in the unfolding process, ensuring that the correction is properly applied.

This unfolding method has been tested for this experiment by using the neutron TOF spectrum produced by the OxyGen simulation (explained in Chap. \ref{chap:OxyGen} to follow) as the input for Bayes' unfolding. Its results replicate the neutron energy spectrum produced by the OxyGen simulation, suggesting that Bayes' method of unfolding is suitable for this experiment.

\section{Solid angle correction}
\label{sec:solidAngleCorrection}

 \begin{figure}[ht]
    \centering
    \includegraphics[width=0.7\textwidth]{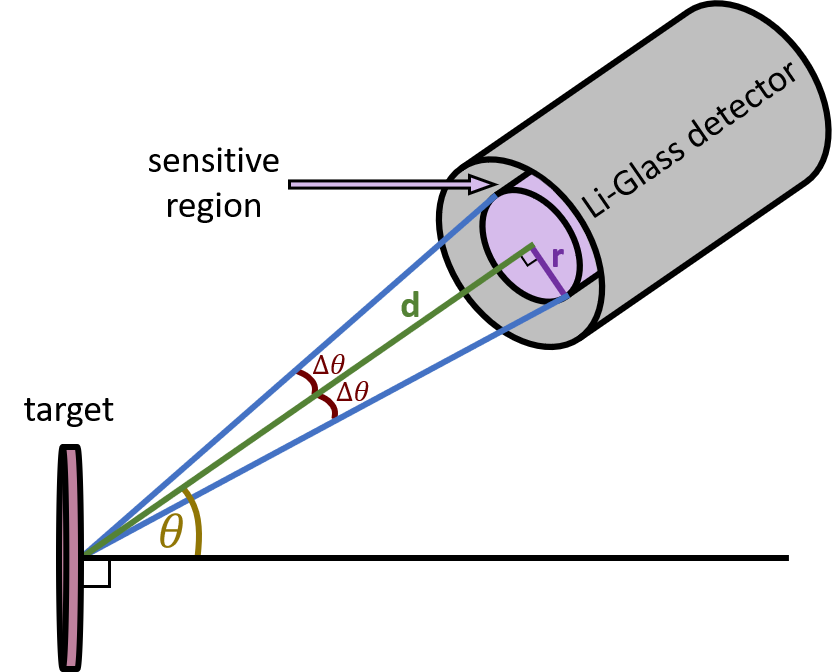}
    \caption{Schematic representation of the solid angle calculation for the Li-glass detector. The detector's sensitive region, corresponding to the $^6$Li component (purple), is positioned at a distance d (green) from the target. The radius of the sensitive region is denoted by r (purple). The detector's angular position relative to the target is given by $\theta$ (gold), while $\Delta\theta$ represents the angular span subtended by the sensitive region. The blue line marks the trajectory from the center of the target to the edge of the sensitive region.}
    \label{fig:solidAngle}
\end{figure}

Following the conversion of TOF spectrum to energy via Bayes' method of unfolding, a solid angle correction was applied. This is due to the fact that the measurement of neutrons occurs only on the horizontal plane where $\phi=0$, while in reality neutrons are emitted in a 3D manner, $0^{\circ} < \phi < 360^{\circ}$. This correction is purely geometric, and it is based on the ratio between the solid angle of the detector and the full solid angle at each polar angle $\theta$ ($0^{\circ}$ to $60^{\circ}$ in jumps of $10^{\circ}$). Only the sensitive region of the detector is accounted for ($^6$Li, the blue part of sketches \ref{fig:geometryFromPTB} and \ref{fig:G4geo} named Li-glass, appearing also in Table \ref{tab:materialDescriptionPTB} and in Fig. \ref{fig:solidAngle}, and the purple component in Fig. \ref{fig:solidAngle}), as only this region involves the actual detection of neutrons.

Generally, the solid angle is defined by:

\begin{equation}
\Omega = \int\int\sin\theta d\theta d\phi
\label{eq:solidAngleDefinition}
\end{equation}

Using the notation presented in Fig. \ref{fig:solidAngle} and plugging it into equation \ref{eq:solidAngleDefinition}, we get:

\begin{equation}
\Omega_{\theta} = \int_0^{2\pi}\int_{\theta-\Delta\theta}^{\theta+\Delta\theta}\sin\theta' d\theta' d\phi = 2\pi[\cos \theta']_{\theta+\Delta\theta}^{\theta-\Delta\theta} = 2\pi[\cos (\theta-\Delta\theta)-\cos (\theta+\Delta\theta)]
\label{eq:solidAnglePluggingIn}
\end{equation}

This solid angle $\Omega_{\theta}$ in equation \ref{eq:solidAnglePluggingIn} must be divided by the solid angle that is detected by the detector. By definition of steradian, the solid angle from the sensitive region of the detector is:

\begin{equation}
\Omega_{det} = \frac{\pi r^2}{d^2}
\label{eq:solidAngleDetector}
\end{equation}

Therefore, the total correction for the solid angle is:

\begin{equation}
f_{\Omega_\theta} = \frac{\Omega_{\theta}}{\Omega_{det}}=\frac{2\pi[\cos (\theta-\Delta\theta)-\cos (\theta+\Delta\theta)]}{\frac{\pi r^2}{d^2}}=2 \left( \frac{d}{r} \right)^2 \left[\cos (\theta-\Delta\theta)-\cos (\theta+\Delta\theta)\right]
\label{eq:solidAngleCorrectionTotal}
\end{equation}

And for $\theta=0^{\circ}$:
\begin{equation}
f_{\Omega_{\theta=0^{\circ}}} =2 \left( \frac{d}{r} \right)^2 \left[1-\cos (\Delta\theta)\right]
\end{equation}

As can be seen in equation \ref{eq:solidAngleCorrectionTotal}, the higher the angle $\theta$ the greater the solid angle correction $f_{\Omega_{\theta}}$. Therefore, upon integrating all angles together in the neutron energy spectrum, the higher scattering angles have a greater effect than the lower ones, due to the solid angle correction.

\chapter{OxyGen}
\label{chap:OxyGen}

The OxyGen simulation can be considered a re-implementation of SimLiT \cite{SimLiT}, as both simulations perform similar functions, with OxyGen focusing on $^{18}$O and SimLiT on $^6$Li. OxyGen is a tool for calculating the neutron energy spectrum from the $^{18}$O$(p,n)^{18}$F reaction with proton energies close to and above the threshold, $E_p=2575-2600$ keV. The OxyGen code is written as a C++ class suitable to use either as a stand-alone tool or for natural incorporation into the GEANT4 transport simulation. OxyGen follows the SimLiT algorithm with a few adaptations.

Similar to SimLiT, OxyGen is a Monte Carlo simulation designed to model the behavior of a single proton. Initially, the energy of the proton is randomly selected according to the beam's energy distribution. The simulation then determines the energy at which a (p,n) reaction occurs within the proton's energy range. Using the kinematics of the reaction, the velocity vector of the emitted neutron in the center of mass (c.m.) frame is calculated. Subsequently, the velocity $V_{cm}$ of the c.m. frame is added to the previously calculated neutron velocity to obtain the velocity vector of the neutron in the laboratory frame (see Fig. \ref{fig:OpnFreaction}). For the purpose of these calculations, the proton beam direction is defined as the positive z-axis.

\section{Calculation details}
\label{sec:howOxyGenWorks}

\begin{figure}[ht]
    \centering
    \includegraphics[width=0.6\textwidth]{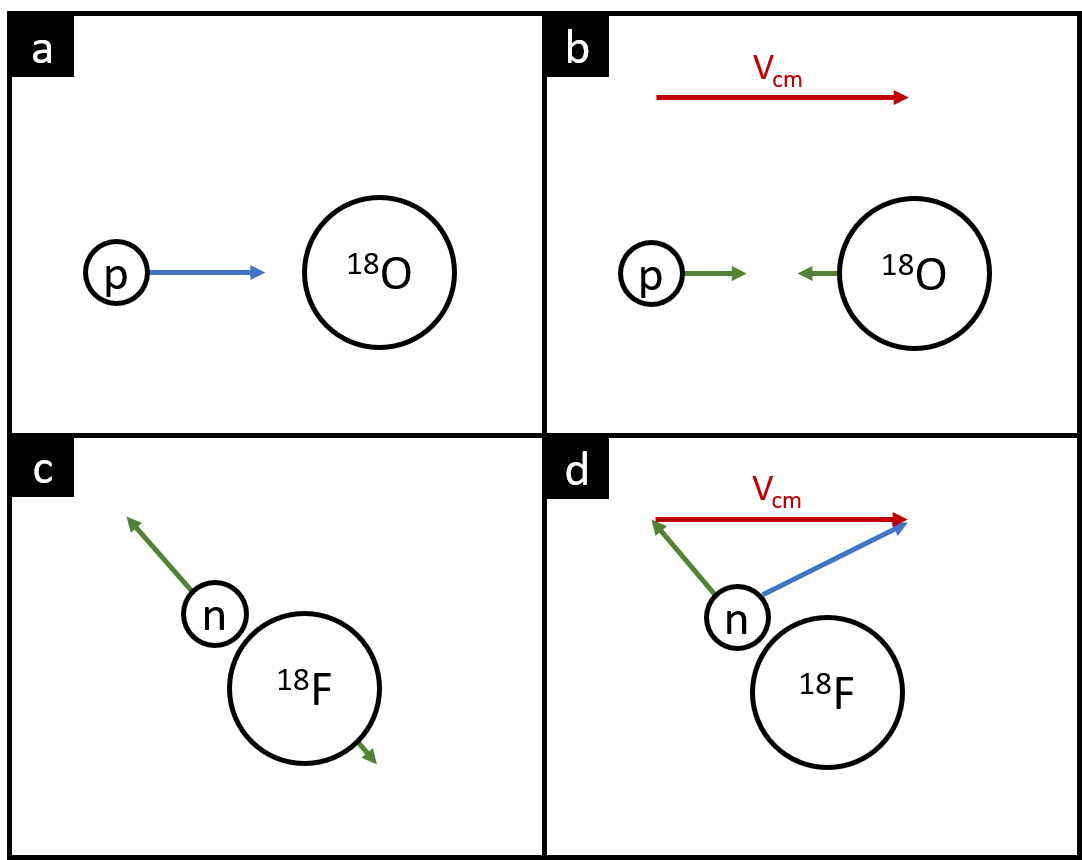}
    \caption{Transformation procedure. (a) In the lab system, the $^{18}$O nucleus is at rest, while the proton has a velocity in z direction (blue arrow). (b) Same picture in the c.m. system. Both $^{18}$O and proton are moving towards each other (green arrows), while the whole system has a velocity $V_{cm}$ in the lab system (red arrow). (c) After the reaction in the c.m. system, a neutron is emitted at a given angle according to the differential cross-section. (d) Same picture in lab system. The velocity of the neutron in the lab (blue) is obtained by adding the velocity of the c.m. system in the lab (red) to the neutron velocity in the c.m. system (green). Figure adopted with minor changes from \cite{SimLiT}.}
    \label{fig:OpnFreaction}
\end{figure}

As previously mentioned, the first step in the OxyGen simulation involves sampling the energy of a proton. For simplicity, OxyGen assumes that the proton beam energy follows a normal distribution, $E_p\sim N(\mu_E,\sigma_E)$, where N represents the normal distribution function, and $\mu_E$ and $\sigma_E$ are user-defined parameters corresponding to the mean energy and standard deviation of the proton beam energy, respectively. The simulation specifically models the $^{18}$O$(p,n)^{18}$F reaction and incorporates the relevant stopping power, which is defined as:

\begin{equation}
F_{stop}(E_p)\equiv-\frac{dE_p}{dz(E_p)}
\label{eq:stoppingPower}
\end{equation}

$F_{stop}$ represents the force exerted by the material on the proton to halt its motion, and its units are $\frac{\text{keV}}{\text{cm}}$. As the proton travels through the material, it loses energy with increasing distance, causing its velocity to decrease until it eventually comes to rest. This energy loss primarily arises from electric interactions between the proton and the electrons in the material. $F_{stop}$ quantifies the rate at which protons decelerate, expressed as a force - the effective force applied by the material to slow the proton. The value of $F_{stop}$ for OxyGen was extracted from SRIM \cite{Ziegler2010}. 

The probability for a (p,n) reaction in the Ta$_2$$^{18}$O$_5$ target is given by:

\begin{equation}
P(E_p)=\frac{f_{^{18}O}\cdot N_A\cdot \rho}{M_A}\cdot 2\pi \int _{0} ^{z_{max}} \int _{0} ^{\pi} \frac{d\sigma}{d\Omega}(E_p(z),\theta_n)\cdot \sin{\theta_n}\cdot d \theta_n \cdot dz
\label{eq:probabilityPNreaction_dz}
\end{equation}

Here, $f_{^{18}O}$ is the percentage of $^{18}$O in the target, $N_A$ is Avogadro's number, $\rho$ is the density of the target in $\frac{g}{cm^3}$ and $M_A$ is the atomic mass of $^{18}$O. The first term in equation \ref{eq:probabilityPNreaction_dz}, $\frac{f_{^{18}O}\cdot N_A\cdot \rho}{M_A}$, represents the number of $^{18}$O atoms per unit volume.
%, essentially describing the amount of $^{18}$O there is in the target.
The remainder of the equation involves a triple integral over the variables $z,\theta$ and $\phi$, where $\theta$ and $\phi$ follow the standard definitions of spherical coordinates. The integration over $\phi$ is represented by the factor $2\pi$, which appears outside the integral in the equation. The variable z represents the distance the proton travels within the target, and the corresponding integral over z accounts for the proton's trajectory. $z_{max}$ is the range of the proton within the target, which corresponds to the target thickness or, in the case where the target is thick enough, the distance before the proton slows down below the reaction threshold. 

The first term in the integral, $\frac{d\sigma}{d\Omega}(E_p(z),\theta_n)$, is the differential cross-section as a function of the proton's energy $E_p(z)$ and the neutron scattering angle $\theta_n$ in the c.m. frame. This term describes the probability of the proton to undergo a reaction for a given energy $E_p(z)$ and neutron scattering angle $\theta_n$. As previously noted, the proton's energy is dependent on the distance it has traveled in the material before reacting, which is why $E_p$ is expressed as a function of z. The remaining part of the integral accounts for the total probability, integrating over all possible angles. These angles are also the neutron scattering angles in the c.m. frame. Therefore, this integral on the cross-section represents the probability of the proton to interact with a single $^{18}$O. Upon multiplying this integral with the first term in equation \ref{eq:probabilityPNreaction_dz}, which represents the amount of $^{18}$O the proton "sees" in the target, we get the probability of the reaction to occur.

Using the definition of $F_{stop}$ in equation \ref{eq:stoppingPower} to change parameters, we get:

\begin{equation}
P(E_p)=\frac{f_{^{18}O}\cdot N_A\cdot \rho}{M_A}\cdot 2\pi \int _{E_{min}} ^{E_p} \int _{0} ^{\pi} \frac{d\sigma}{d\Omega}(E_p,\theta_n)\cdot \frac{1}{F_{stop}(E_p)}\cdot \sin{\theta_n}\cdot d \theta_n \cdot dE_p
\label{eq:probabilityPNreaction_dEp}
\end{equation}

Once it is determined whether or not the proton has interacted, the next step is to identify the energy of the proton upon reaction in the lab frame and in the c.m. frame, and to calculate the velocity of the c.m. frame in the lab, $V_{cm}$. A random emission angle is assigned to the neutron based on Legendre polynomials explained in Sec. \ref{subsec:P-wave}. A random energy is assigned to the proton based on the integrand of \ref{eq:probabilityPNreaction_dEp} .

The velocity vector of the neutron in the c.m. system can then be calculated by using the Q-value of the $^{18}$O$(p,n)^{18}$F reaction, $Q=-2438.3$ keV, and conservation of energy and momentum by:

\begin{equation}
E_{n,cm}=\frac{M_{^{18}F}}{M_n+M_{^{18}F}}\cdot(E_{p,cm}+Q)
\label{eq:En-cm}
\end{equation}

The velocity of the neutron in the laboratory system can then be calculated by adding the neutron's velocity in the c.m. system to $V_{cm}$, as illustrated in Fig. \ref{fig:OpnFreaction}.

\subsection{$^{18}$O$(p,n)^{18}$F cross-section}
\label{subsec:CSbairCalculation}

\begin{figure}[ht]
    \centering
    \includegraphics[width=0.6\textwidth]{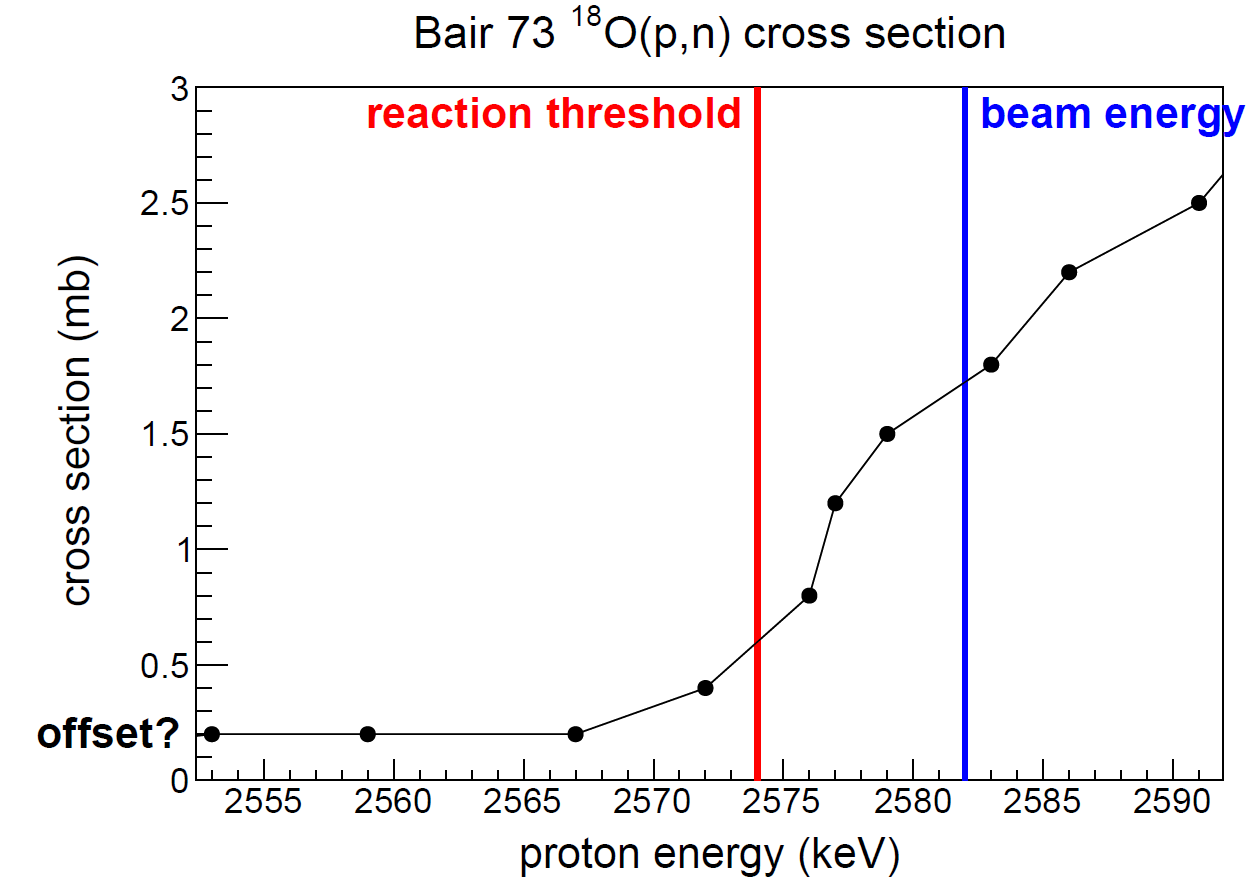}
    \caption{$^{18}$O$(p,n)^{18}$F reaction near-threshold cross-section measured by Bair \cite{Bair}. The red vertical line is the reaction threshold, and the blue vertical line is the $E_p=2582$ keV energy used by Heil \textit{et al.} \cite{Heil2005}. The tabulated data is taken from the EXFOR database \cite{Otuka2014}. There is a 0.2 mb offset in the cross-section.}
    \label{fig:BairCS}
\end{figure}

\begin{figure}[ht]
    \centering
    \includegraphics[width=0.6\textwidth]{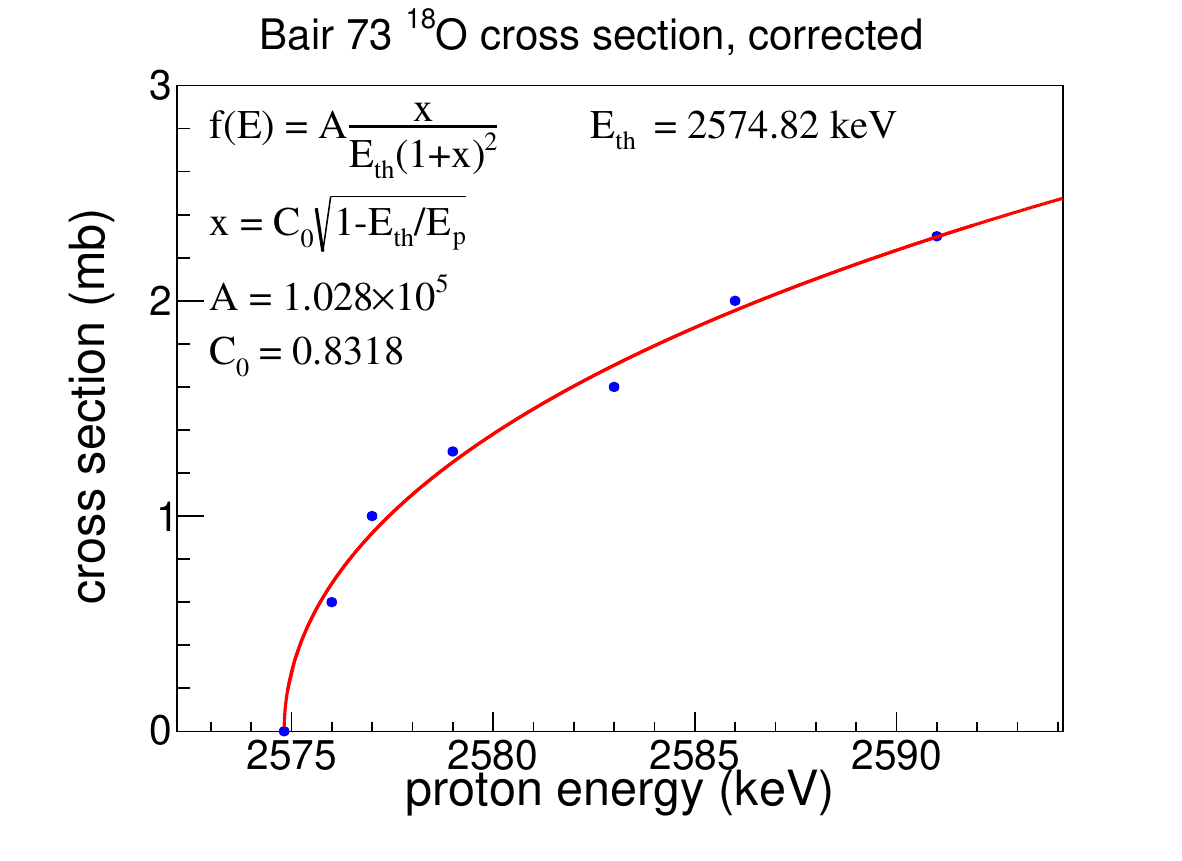}
    \caption{A corrected excitation curve, after removing the 0.2 mb offset, fitted to the form suggested by Lee and Zhou \cite{Lee1999}.}
    \label{fig:BairCSfit}
\end{figure}

The $^{18}$O$(p,n)^{18}$F cross-section is taken from experimental data. The only measurement of the cross-section which is reasonably detailed and precise close to the reaction threshold was published by J. K. Bair \cite{Bair} (see Fig. \ref{fig:BairCS}). 
%Other available measurements are either too sparse \cite{Hess2001HE34} or suffer from large uncertainties \cite{Ruth1979} close to the reaction threshold. 
As noted by Lee and Zhou \cite{Lee1999}, the yields of forward neutrons in the laboratory reference frame result from a competition between a vanishing term and an infinite term, leading to irregular behavior when using tabulated data to calculate the neutron energy spectrum. Therefore, their recommended excitation function was used, which is:
\begin{equation}
    \frac{d\sigma}{d\Omega^{\prime}}=\frac{A}{4\pi}\frac{x}{E_p\left(1+x\right)^2}
\label{eq:Lee}
\end{equation}
$E_p$ is the proton energy, $x=C_0\sqrt{1-E_{th}/E_p}$, $E_{th}$ the reaction energy threshold, and $C_0, A$ are constants to be determined. $d\sigma/d\Omega^{\prime}$ is the differential cross-section in the c.m. frame as a function of the energies in the laboratory frame. Eq. \ref{eq:Lee} was fitted to the cross-section data from Bair's data \cite{Bair} to determine $C_0$ and $A$ (see Fig. \ref{fig:BairCSfit}). For the fit, we forced the cross-section to be zero at the reaction threshold. This required removing a clear offset of 0.2 mb in the tabulated data downloaded from EXFOR \cite{Otuka2014}. The use of Eq. \ref{eq:Lee}, despite improving the situation in forward angles, does not completely remove the singularity and we can still see an unphysical "bump" at $0^{\circ}$ around 7.5 keV, which corresponds to zero energy in the c.m. frame.

\subsection{P-wave component}
\label{subsec:P-wave}

Eq. \ref{eq:Lee} reflects a pure s-wave in the c.m. frame. This assumption is not necessarily true. Given the proton radius $R_p = 0.84$ fm \cite{Tiesinga2021} and the $^{18}$O nucleus radius of $R_{^{18}O}=2.77$ fm \cite{Angeli2013}, along with the reduced de Broglie wavelength of the incident proton at $E_p = 2580.5$ keV ($\bar\lambda = 2.84$ fm), a non-negligible p-wave component cannot be excluded. As noted in Krane \cite{Krane}, the ratio $\frac{R}{\bar\lambda}$, which is approximately 1.25 in our case, can provide an estimate on the maximum partial wave expected to be involved in the interaction. Therefore, a $l=1$ contribution cannot be ruled out, and a p-wave component in the c.m. frame was allowed in the simulation. Since no experimental differential cross-sections are available, we used our experimental data to evaluate the p-wave component. 

This is manifested by the angle of the neutron randomly chosen using an expansion in Legendre polynomials \cite{Blatt1952}:
\begin{equation}
    \left(\frac{d\sigma}{d\Omega}\right)_{cm} \propto \sum_{l=0}^{l=2}A_lP_l(\cos\theta ).
\end{equation}
The code can be set to use Legendre polynomials up to $l=2$, although we found a better fit to the data by using only $l=0,1$. 

\section{OxyGen as input to Geant4 simulation}
\label{sec:OxyGenGeant4}

As mentioned earlier, the result of the OxyGen simulation is a neutron energy spectrum. The Geant4 transport simulation described in Sec. \ref{sec:G4} uses a neutron energy distribution as an input (see Sec. \ref{subsec:inputToG4}). Using a uniform distribution of neutrons between (0,30) keV has the advantage of providing a response matrix without prior bias. However, using OxyGen as an input provides a direct comparison between calculated and experimental TOF spectra, avoiding systematic inaccuracies in the unfolding method. It is important to highlight that when this is done, the tantalum backing of the $^{18}$O target is also taken into account as part of the geometrical components of the Geant4 simulation, as shown in Fig. \ref{fig:G4geo}.

\begin{figure}[ht]
    \centering
    \includegraphics[width=\textwidth]{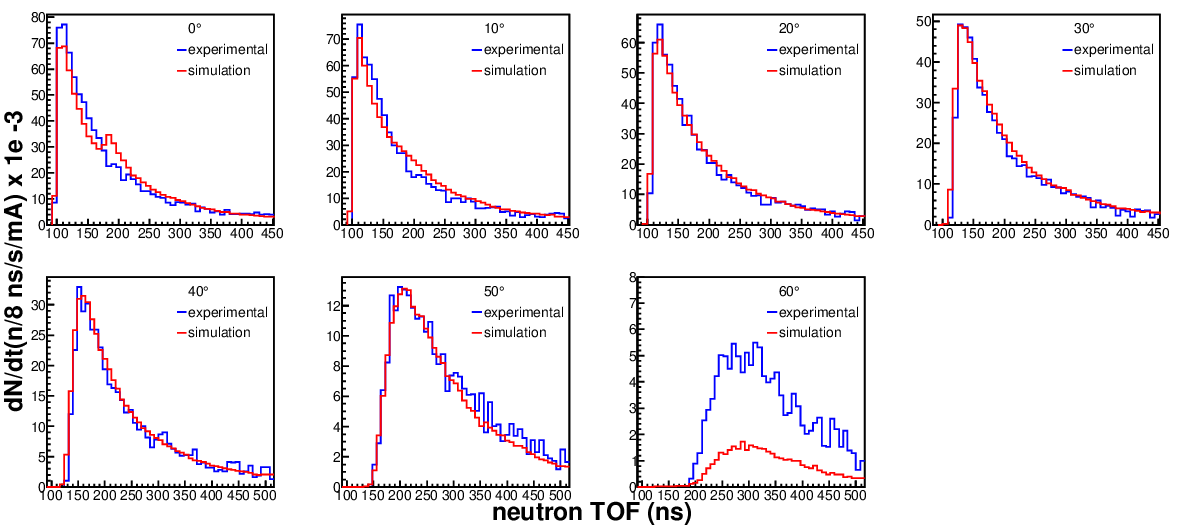}
    \caption{These histograms show the comparison between experimental and simulated time of flight (TOF) spectrum for all angles. The simulated TOF spectrum for each angle is normalized to that of the 0\textdegree\ angle.}
    \label{fig:ExpSimTOF_allAngles}
\end{figure}

Using OxyGen as the source for the Geant4 simulation, the result of the TOF spectrum produced by the simulation can be compared with the experimental TOF spectrum for each angle individually, as shown in Fig. \ref{fig:ExpSimTOF_allAngles}. A single normalization of the simulated TOF histograms in all angles was used to match the TOF integral at 0\textdegree. The slight peak observed in the TOF spectrum for 0\textdegree\ for the simulation at ~180 ns result from the competition between the vanishing term and infinite term as explained by Lee and Zhou \cite{Lee1999} and mentioned in Sec. \ref{subsec:CSbairCalculation}.

It should be noted that the experimental TOF data were used in order to tune the Legendre polynomial coefficients to be $A_0=1.4, A_1=-0.4$ , hence, those two datasets are not completely independent. An independent comparison at different proton beam energies will help to validate those values. The disagreement at 60\textdegree\ will be discussed in Chapter \ref{chap:results_and_discussion}.

\subsection{Determining the proton beam energy}\label{subsec:ProtonEnergies}

To determine the proton beam energy with higher precision than that provided by the standard threshold scan performed by the beam operators, we used the OxyGen simulation as an input to the Geant4 transport simulation. By varying the proton beam energy in OxyGen, we generated corresponding neutron energy spectra, which were then used in Geant4 to calculate TOF spectra at different detection angles. These simulated TOF spectra were compared directly with the experimental TOF histograms.

The endpoint of the TOF spectrum is determined purely by reaction kinematics and is therefore independent of model dependent quantities such as the cross-section, angular distribution or stopping power used in OxyGen. This makes the TOF endpoint a reliable observable for determining the true proton beam energy.

\begin{figure}[ht]
    \centering
    \includegraphics[width=\textwidth]{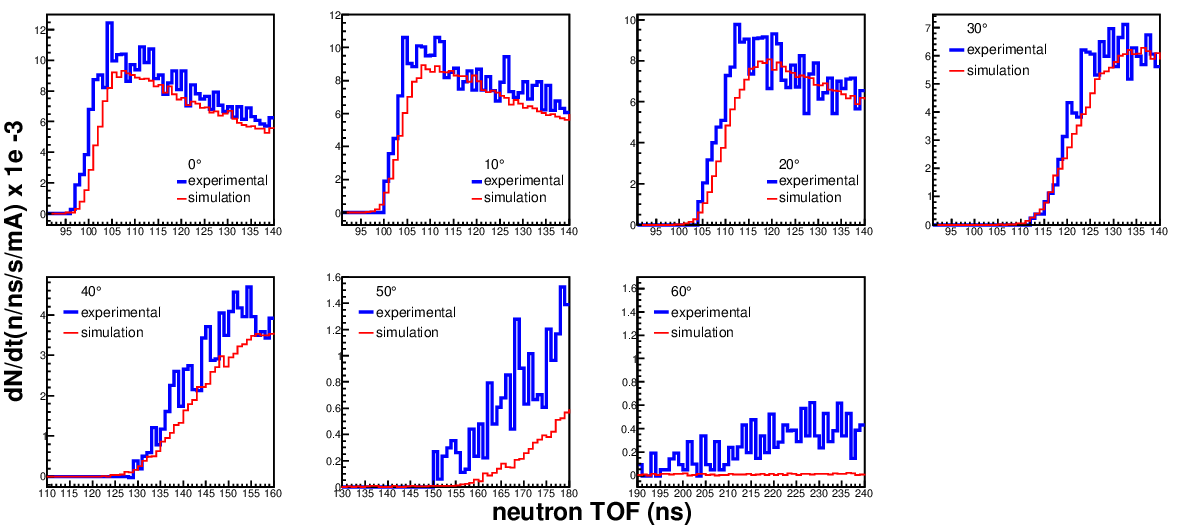}
    \caption{These histograms show a comparison between experimental and simulated TOF spectra for all angles, focusing on the TOF endpoint. The simulated TOF spectrum for each angle is normalized to that of the 0\textdegree\ angle. The simulated TOF spectrum for each angle was calculated using OxyGen with the proton energy of $2580.0$ keV.}
    \label{fig:ExpSimTOF_allAngles2580zoomed}
\end{figure}

\begin{figure}[ht]
    \centering
    \includegraphics[width=\textwidth]{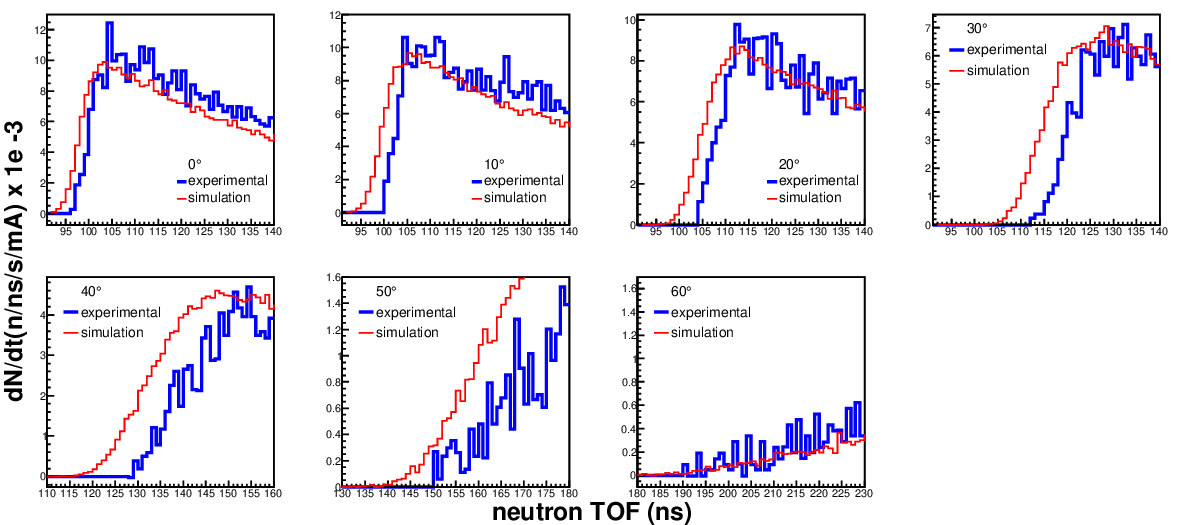}
    \caption{These histograms show a comparison between experimental and simulated TOF spectra for all angles, focusing on the TOF endpoint. The simulated TOF spectrum for each angle is normalized to that of the 0\textdegree\ angle. The simulated TOF spectrum for each angle was calculated using OxyGen with the proton energy of $2581.0$ keV.}
    \label{fig:ExpSimTOF_allAngles2581zoomed}
\end{figure}

\begin{figure}[ht]
    \centering
    \includegraphics[width=\textwidth]{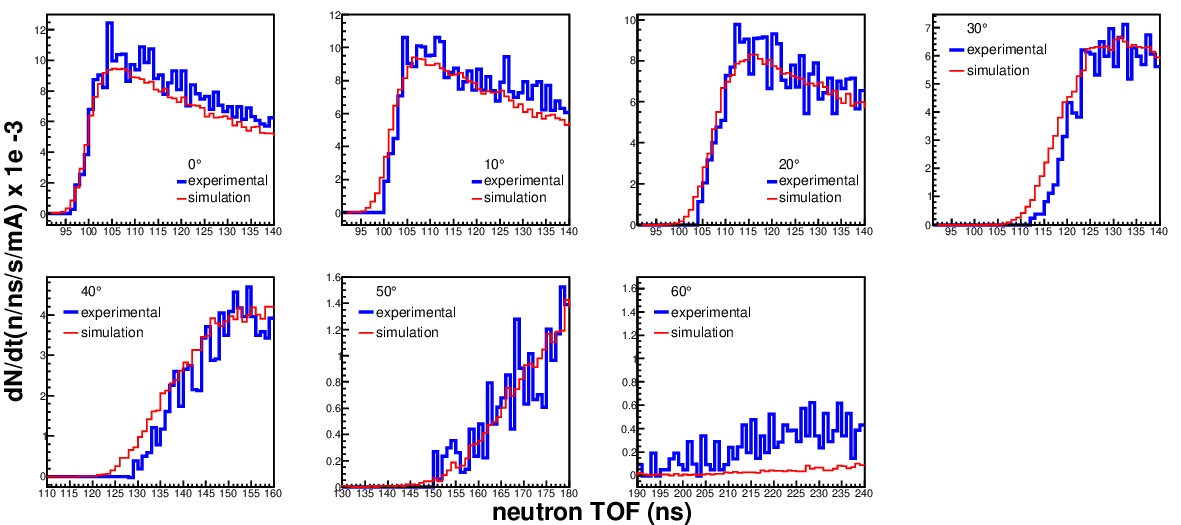}
    \caption{These histograms show a comparison between experimental and simulated TOF spectra for all angles, focusing on the TOF endpoint. The simulated TOF spectrum for each angle is normalized to that of the 0\textdegree\ angle. The simulated TOF spectrum for each angle was calculated using OxyGen with the proton energy of $2580.5$ keV.}
    \label{fig:ExpSimTOF_allAngles2580.5zoomed}
\end{figure}

Figures \ref{fig:ExpSimTOF_allAngles2580zoomed} - \ref{fig:ExpSimTOF_allAngles2580.5zoomed} show the comparison between experimental and simulated TOF spectra for proton energies of 2580.0, 2580.5, and 2581.0 keV. In the 2580.0 keV simulation, the simulated TOF endpoint appears shifted too far to the right, meaning that the simulation predicts neutrons arriving later than those observed experimentally. Since neutrons cannot arrive earlier than allowed by kinematics, this energy must be too low. Conversely, the 2581.0 keV simulation shows the TOF endpoint shifted too far to the left, predicting earlier arrivals than observed, even for the neutrons detected (in the simulation) at the front of the sensitive region without any neutron scattering. The best agreement is obtained for a proton energy of 2580.5 keV, as shown in Figs. \ref{fig:ExpSimTOF_allAngles} and \ref{fig:ExpSimTOF_allAngles2580.5zoomed}, where the simulated and experimental TOF endpoints coincide across all angles.

Based on this comparison, the proton beam energy was determined to be $E_p=2580.5\pm0.2$ keV. This value was adopted for all subsequent analyses.

\chapter{Results and Discussion}
\label{chap:results_and_discussion}

 \begin{figure}[ht]
    \centering
    \includegraphics[width=\textwidth]{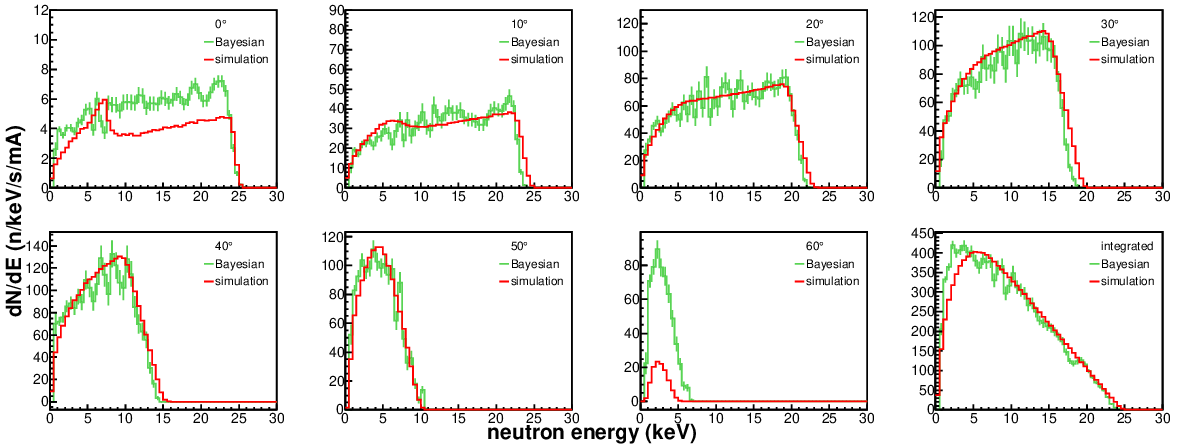}
    \caption{Experimental neutron spectra (green) compared to OxyGen (red) at each angle and the angle integrated spectrum. The experimental data were obtained from the TOF neutron spectra using Bayes' unfolding method. The error bars represents statistical uncertainties only. The simulated data is normalized to the total experimental yields between 0\textdegree\ and 50\textdegree.}
    \label{fig:energyComparison_allAngles}
\end{figure}

Fig. \ref{fig:energyComparison_allAngles} shows the extracted neutron energy spectra at each angle as well as the integrated spectrum. The data are compared to OxyGen and normalized to the integrated yields between 0\textdegree\ and 50\textdegree\ to avoid the clear disagreement at 60\textdegree. Fig \ref{fig:ExpSimTOF_allAngles} compares between the experimental TOF data and the OxyGen+Geant4 simulated data.

Both comparisons show good agreement between 10\textdegree\ and 50\textdegree, and there is also a reasonable agreement between the 0\textdegree\ TOF spectra. The experimental integrated spectrum has an excess of events below 5 keV compared to OxyGen, which is a result of the disagreement at 60\textdegree. These disagreements will be discussed in the following sections.

\section{Underestimation at 0\textdegree}

\begin{figure}[ht]
    \centering
    \includegraphics[width=\textwidth]{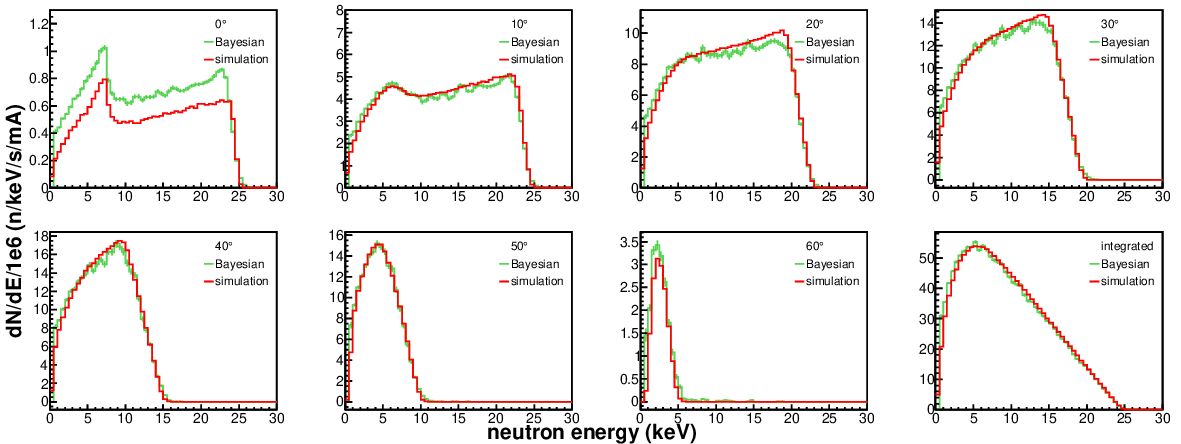}
    \caption{Simulated energy spectrum converted from the simulated TOF spectrum by Bayes' unfolding method, as well as a comparison to the simulated energy spectrum directly from OxyGen. This is presented for all angles separately and integrated together.}
    \label{fig:simAsInput}
\end{figure}

The underestimation observed at 0\textdegree\ is specifically present when converting to the neutron energy spectrum, but does not manifest in the neutron TOF spectrum itself (see Fig. \ref{fig:ExpSimTOF_allAngles}). To identify the source of this issue, a test was performed by applying Bayes' unfolding method (see Sec. \ref{sec:BayesUnfoldingMethod}) onto a "response matrix" generated by the Geant4 simulation using OxyGen as input (see Sec. \ref{sec:OxyGenGeant4}). This approach converted the simulated TOF spectrum to the energy spectrum, which was then compared to the neutron energy spectrum generated by the OxyGen simulation directly. This is presented in Fig. \ref{fig:simAsInput}. Bayes' unfolding spectrum closely matched the direct OxyGen energy spectrum, except at 0\textdegree\ where the direct OxyGen energy spectrum underestimated the neutron energy spectrum obtained from Bayes' unfolding method.

To investigate further, simulations were conducted at various angles focusing only on the detector's sensitive region, and compared to simulations that accounted for the entire neutron spectrum as generated by OxyGen. Upon comparison, it was found that the TOF spectrum contains neutrons that did not solely originate from the sensitive region. While this effect was noticeable at all angles, it was most pronounced at 0\textdegree\ where the discrepancy between the expected and detected neutron trajectories was most evident.

These simulations suggest that neutron scattering is most pronounced at 0\textdegree. The reason that it doesn't appear in the neutron TOF spectrum (Fig. \ref{fig:ExpSimTOF_allAngles}), is because this scattering is taken into account by the Geant4 simulation. The simulation used to compare to the experimental data in Fig. \ref{fig:ExpSimTOF_allAngles} is the OxyGen simulation with Geant4.

One question that arises is why neutron scattering is more significant at 0\textdegree\ compared to other angles. At 0\textdegree\ there are fewer neutrons emitted from the $^{18}$O$(p,n)^{18}$F reaction. As shown in Fig. \ref{fig:energyComparison_allAngles}, fewer neutrons are detected at 0\textdegree\ than at other angles. This can be attributed to the small solid angle at 0\textdegree. 

From this perspective, the source of neutrons arriving at any given angle can be seen as a combination of the primary neutron source and scattered neutrons. The low neutron count at 0\textdegree\ amplifies the contribution of scattered neutrons, effectively increasing the background at this angle. However, due to the inherent limitations of the detector geometry and the difficulty of accounting for this additional background without introducing bias, this effect remains unaccounted for in the analysis. In principle this can be fixed by calculating a response matrix using OxyGen as an input instead of a uniform and isotropic energy distribution, but this will introduce a major bias to the analysis.

% Alternatively, the increased prominence of neutron scattering at 0\textdegree\ can be attributed to the velocity dependence of the scattering probability. Neutrons emitted at 0\textdegree\ are generally faster than those emitted at other angles. The probability of neutron scattering is inversely proportional to the neutron velocity (i.e., $\frac{1}{v}$, where $v$ is the velocity), meaning that neutrons with lower velocities are more likely to scatter. Consequently, neutrons emitted at angles other than 0\textdegree, which have lower velocities, are more prone to scattering. This can result in a higher number of scattered neutrons being detected at 0\textdegree\ compared to other angles, even if the scattering is uniformly distributed across the remaining angles.

It is important to note that although this underestimation at 0\textdegree\ is noticeable in the individual neutron energy spectrum for 0\textdegree, its impact on the integrated spectrum is minimal. The number of neutrons detected at 0\textdegree\ is small relative to other angles, and the correction for the solid angle is minor as well. As explained in Sec. \ref{sec:solidAngleCorrection}, the correction for the solid angle decreases with smaller angles, further reducing the effect on the integrated spectrum. Therefore, while the underestimation at 0\textdegree\ is observable in the individual energy spectrum, it does not significantly influence the integrated spectrum.

In addition to the underestimation, there is a unique peak-like structure around $\sim7$ keV at forward directions. This structure is a numerical artifact that was previously discussed in literature \cite{Lee1999,SimLiT}, and its effect on the total yields is very low.

\section{Underestimation at 60\textdegree}

The underestimation observed at $60^\circ$ appears both in the neutron TOF spectrum (Fig.~\ref{fig:ExpSimTOF_allAngles}) and in the unfolded neutron energy spectrum. It is therefore natural to analyze the source of the discrepancy in the TOF domain, since the energy spectrum is derived directly from the TOF data and the unfolding procedure has been validated independently, as demonstrated by the successful unfolding of simulated data (Fig.~\ref{fig:simAsInput}). The fact that the discrepancy has the same magnitude in both domains also indicates that the solid-angle correction is not responsible. Several possible explanations for the observed discrepancy are discussed below.

\subsubsection*{Inaccuracy in OxyGen}
A first possibility is that OxyGen underestimates the neutron yield at $60^\circ$. Although this cannot be ruled out, the good agreement at all other angles suggests that a major error in the kinematic treatment is unlikely. Including a small $d$-wave component in the Legendre expansion does not significantly increase the predicted yield at this angle, and larger modifications of the angular distribution would degrade the agreement at the dominant forward angles. Nevertheless, the angular distribution at large angles is inherently less constrained, and this possibility will be further addressed below.

\subsubsection*{Effect of proton energy}
However, we can be reasonably confident that the underestimation of neutrons at $60^\circ$ is not caused by a slightly higher proton energy. While such a shift could, in principle, improve the agreement between the experimental and simulated energy spectra, a direct comparison in the TOF domain contradicts this scenario. As discussed in Sec.~\ref{subsec:ProtonEnergies}, for both higher and lower assumed proton energies, the comparison between experiment and simulation shows clear discrepancies in the arrival times of the first neutrons, making this explanation physically unlikely. In particular, for higher proton energies, Fig.~\ref{fig:ExpSimTOF_allAngles2581zoomed} demonstrates that the earliest arriving neutrons disagree with the measured TOF spectrum, indicating that an incorrect proton energy cannot account for the discrepancy observed at $60^\circ$.

\subsubsection*{Background subtraction}
Because the absolute neutron yield is lowest at $60^\circ$, the signal-to-background ratio is poorest at this angle, and an underestimation of the background level would artificially enhance the measured yield. The uncertainties presented in this work include only statistical contributions and do not account for such systematic effects. Still, even conservative estimates of the background uncertainty at $60^\circ$ remain too small to explain the magnitude of the discrepancy, and therefore background subtraction alone cannot account for the observed difference.

\subsubsection*{Angular misalignment}
Misalignment of the detector angle could, in principle, influence the measured yield, especially since the angular distribution decreases rapidly at large angles. While no independent measurement of the rail markers exists, simulations indicate that small deviations of $1^\circ$--$2^\circ$ are insufficient on their own to reproduce the full discrepancy. Nonetheless, this effect cannot be fully excluded without further study and will be further addressed below.

\subsubsection*{Neutron scattering}
The Geant4 model includes neutron scattering in the target backing and detector assembly, but does not capture all possible scattering sources in the experimental hall. At $60^\circ$, where the direct neutron flux is weakest, additional scattered neutrons originating from surrounding structures could make a relatively large contribution. Such neutrons would primarily populate the long-TOF (low-energy) region, consistent with the observed pattern. However, no direct evidence is currently available to confirm this hypothesis.

\subsubsection*{Outlook: analysis of the additional dataset}
\begin{figure}[ht]
    \centering
    \includegraphics[width=0.6\textwidth]{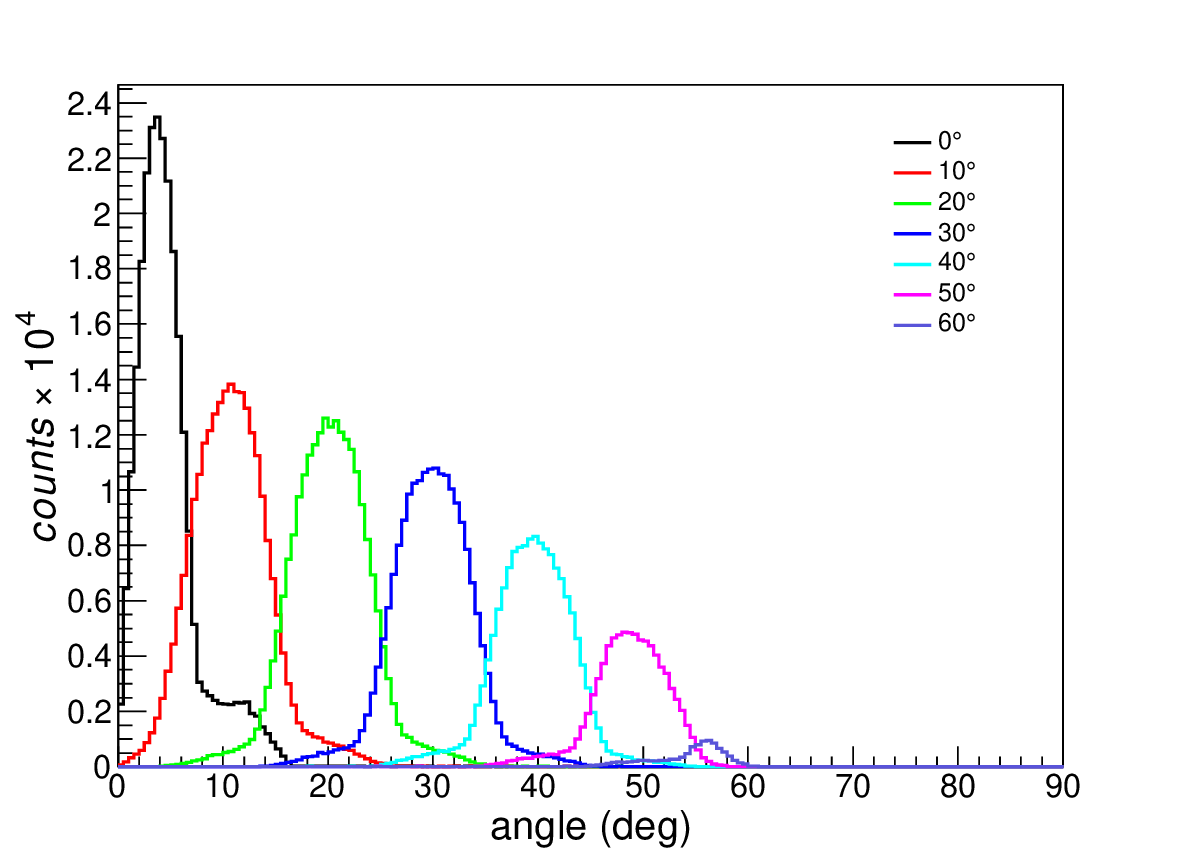}
    \caption{Calculated detector acceptance. The different colors represent the simulated vertex angular distribution detected at each detector angle, extracted using the Geant4 simulation with OxyGen as the neutron source.}
    \label{fig:angleSpectrum}
\end{figure}
For the proton energy used in this experiment, the neutron angular distribution barely reaches 60\textdegree, as shown in Fig. \ref{fig:angleSpectrum}. Furthermore, a substantial fraction of the neutrons detected at 60\textdegree\ are actually neutrons scattered from lower angles, and the vertex angular distribution of these detected neutrons are not centered at 60\textdegree. 

As discussed in Sec.~\ref{sec:ExpHistory}, an additional dataset was recorded at the same facility with the same experimental setup a few months earlier. Although the proton energy in that run was several~keV too high for the main analysis, it is particularly valuable in the context of the $60^\circ$ discrepancy. At higher beam energy, the neutron angular distribution extends beyond $60^\circ$, providing a much improved signal-to-background ratio at this angle. While not all angles were measured in that run, comparison between experiment and OxyGen up to $60^\circ$ at the higher proton energy would allow the angular-distribution model to be better constrained and reduce the sensitivity of the analysis to detector alignment and scattering effects. Unfortunately, such an analysis could not be completed within the scope of this work and is therefore left as a direction for future investigation.

\section{Comparison to Heil \textit{et al.}}

 \begin{figure}[ht]
    \centering
    \includegraphics[width=0.7\textwidth]{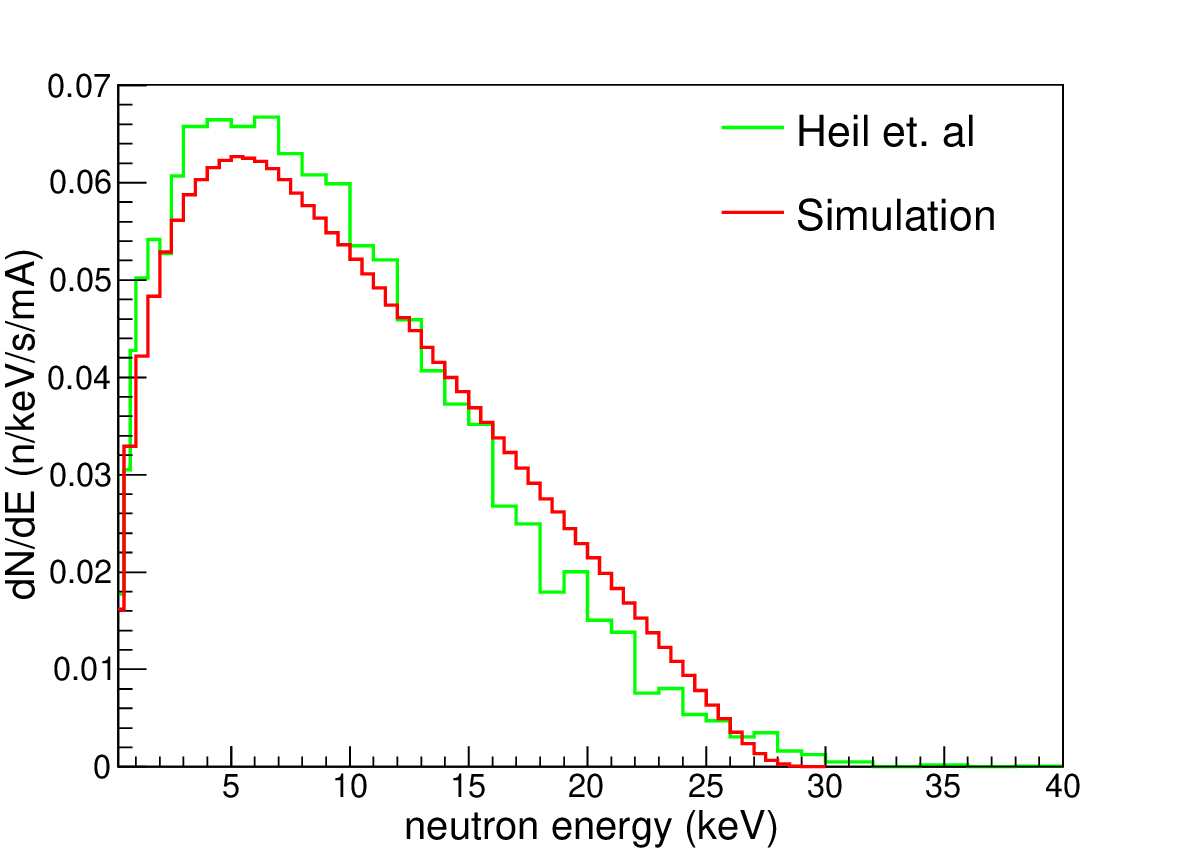}
    \caption{Heil \textit{et al.}'s data compared to the OxyGen simulation. The simulation was done at the same proton energy that Heil \textit{et al.}'s data was measured: 2582 keV.}
    \label{fig:Heil}
\end{figure}

Heil \textit{et al.}'s measurements~\cite{Heil2005} were conducted with a proton energy of $2582$~keV, which differs slightly from the proton energy of $2580.5$~keV used in this experiment. Because of this small discrepancy in proton energy, a direct comparison of the neutron energy spectra from the two experiments is not straightforward. Nevertheless, running the OxyGen simulation at the proton energy of Heil \textit{et al.}'s experiment ($2582$~keV) shows that the difference between the simulated and experimental neutron energy spectra is minimal, as illustrated in Fig.~\ref{fig:Heil}.

Although the integrated spectrum alone does not allow for firm conclusions about the behavior at individual angles, a comparison between the integrated energy spectrum of OxyGen and that of Heil \textit{et al.} does not exhibit the same 0--5~keV discrepancy between experiment and simulation that is observed at higher angles in our data. This observation suggests that the issue seen at $60^\circ$ in our experiment may be related to the low relative neutron yields at this angle for $E_p = 2580.5$~keV, where small systematic effects become comparatively more significant.

\section{Planning future experiments at SARAF}

The OxyGen simulation was created to plan for future experiments at SARAF using the $^{18}$O target as a neutron source, as it is able to determine the energy spectrum of the neutron source.
%In order to plan for future experiments at SARAF using the $^{18}$O target as a neutron source, the OxyGen simulation was created in order to determine the energy spectrum of the neutron source. 
In addition, the Geant4 simulation that uses OxyGen as the neutron source can be used to plan and analyze future experiments at SARAF. Although these simulations are not in perfect agreement to the experimental data, especially at high angles, they are in fairly good agreement both to the experiment we conducted in collaboration with PTB and the experiment conducted by Heil \textit{et al}. 

In conclusion, the OxyGen simulation calculates the energy spectrum of the neutrons reasonably well, and can be used as an important tool for future experiments using the $^{18}$O$(p,n)$ as a quasi-stellar neutron source.

\chapter{Summary}
\label{chap:summary}

This thesis focused on investigating the neutron energy spectrum resulting from the $^{18}$O$(p,n)^{18}$F reaction, with a specific emphasis on comparing experimental data to the OxyGen simulation in preparation for future experiments at SARAF. An experiment was conducted at PTB, where protons with an energy of $2580.5$ keV were projected onto an $^{18}$O target, inducing neutron emission, which was then detected by a $^6$Li-glass detector.

To analyze the experimental data more accurately, we developed a Geant4 transport simulation to convert the neutron TOF spectrum into the neutron energy spectrum using Bayes' unfolding method. The resulting energy spectrum was then compared to the neutron energy spectrum predicted by the OxyGen simulation. The comparison showed good agreement between the experimental data and the simulation. However, at high angles, the simulation underestimated the experimental data, possibly due to a low number of neutrons detected at these angles, compounded by the contribution of scattered neutrons. Analysis of a similar experiment at slightly higher proton energies can determine if this indeed was the cause. 

Compared to a previous experiment conducted by Heil \textit{et al.} \cite{Heil2005}, OxyGen showed similar results, likely with no issues at higher angles, possibly due to the higher proton energy (2582 keV) used in their experiment, which resulted in a higher number of neutrons at higher angles. This suggests that the OxyGen simulation is effective in predicting the neutron energy spectrum for the $^{18}$O$(p,n)^{18}$F reaction at different proton energies.

Additionally, the OxyGen simulation was used as an input for a Geant4 simulation, which generated simulated data for the TOF neutron spectrum. This step was particularly crucial in determining the proton energy used in the experiment, allowing for a more accurate comparison between the experimental and simulated results.

Overall, the findings of this thesis demonstrate that the OxyGen simulation provides a reliable prediction of the neutron energy spectrum for the $^{18}$O$(p,n)^{18}$F reaction and can be effectively used to plan and analyze future experiments at SARAF for different proton energies.

\bibliography{thesis}

\end{document}